\pdfoutput=1
\documentclass[lettersize,journal]{IEEEtran}
\usepackage{hyperref}
\usepackage{url}
\usepackage{subfig}
\usepackage{booktabs}
\usepackage{threeparttable}
\usepackage{adjustbox}
\usepackage{graphicx}
\usepackage[normalem]{ulem}
\usepackage{multirow}
\usepackage{textcomp}
\usepackage{caption}

\usepackage[fleqn]{amsmath}
\usepackage{amsmath,amssymb,amsfonts}
\usepackage{pifont}
\usepackage{algorithmic,algorithm}
\usepackage{mathtools}

\usepackage{todonotes}
\usepackage{lipsum}
\usepackage{xcolor}
\useunder{\uline}{\ul}{}

\let\labelindent\relax
\usepackage{enumitem}

\usepackage[en-US]{datetime2}

\makeatletter
\newcommand{\monthyear}{%
  \DTMenglishmonthname{\@dtm@month}, \@dtm@year
}
\makeatother

\makeatletter
\newcommand{\thisyear}{%
 \@dtm@year
}
\makeatother

\usepackage{bbding}
\usepackage[normalem]{ulem}
\useunder{\uline}{\ul}{}
\usepackage[
    style=ieee,
    citestyle=numeric-comp,
    backend=biber,
    hyperref=true,
    maxnames=1,
    maxbibnames =3,
    minnames=1,
    mincrossrefs=1,
    natbib=true,
    sorting=none,
    sortcites=true,
    url=false,
    doi=false,
    eprint=false
]{biblatex}
\AtEveryBibitem{%
  \clearname{translator}%
  \clearlist{publisher}%
  \clearfield{pagetotal}%
\clearfield{pagetotal}%
\clearfield{issn}%
\clearfield{editor}%
\clearfield{month}%
\clearfield{notes}%
\clearfield{language}%
}
\DeclareFieldFormat
  [article,inbook,incollection,inproceedings,patent,thesis,unpublished]
  {title}{{#1}}

\def\BibTeX{{\rm B\kern-.05em{\sc i\kern-.025em b}\kern-.08em
    T\kern-.1667em\lower.7ex\hbox{E}\kern-.125emX}}
\newcommand{\revise}[1]{\textcolor{black}{#1}}

\definecolor{applegreen}{rgb}{0.55, 0.71, 0.0}
\newcommand{\yesmark}{\checkmark}
\newcommand{\xmark}{\ding{55}}%

\newcommand{\Reals}{\mathbb{R}}

\newcommand{\Loss}{\mathcal{L}}

\def\BibTeX{{\rm B\kern-.05em{\sc i\kern-.025em b}\kern-.08em
    T\kern-.1667em\lower.7ex\hbox{E}\kern-.125emX}}
\begin{document}

\title{Large Language Model-informed ECG Dual Attention Network for Heart Failure Risk Prediction}

\author{Chen Chen,
        Lei Li,
        Marcel Beetz, 
        Abhirup Banerjee, 
        Ramneek Gupta, 
        Vicente Grau
\thanks{Manuscript received \today. 
This work is supported by the Novo Nordisk collaborative research fund.}
\thanks{Chen Chen, Lei Li,
        Marcel Beetz, 
        Abhirup Banerjee, and
        Vicente Grau are with the Institute of Biomedical Engineering, Department of Engineering Science, University of Oxford, Oxford, United Kingdom. Chen Chen is also an Honorary Research Associate at Imperial College London; Lecturer in Computer Vision at University of Sheffield. This work is mainly done during her time at Oxford. Ramneek Gupta is with the Novo Nordisk Research Centre Oxford (NNRCO).
        (Corresponding author: Chen Chen, email: work.cherise@gmail.com)}
\thanks{A. Banerjee is a Royal Society University
Research Fellow and is supported by the Royal Society Grant No.
URF\textbackslash R1\textbackslash221314. The works of A. Banerjee and V. Grau were partially supported by the British Heart Foundation Project under Grant PG/20/21/35082 and by the CompBioMed 2 Centre of Excellence in
Computational Biomedicine (European Commission Horizon
2020 research and innovation programme, grant agreement
No. 823712).}
}

\maketitle  
\begin{abstract}

\revise{Heart failure (HF) poses a significant public health challenge, with a rising global mortality rate. Early detection and prevention of HF could significantly reduce its impact. We introduce a novel methodology for predicting HF risk using 12-lead electrocardiograms (ECGs). We present a novel, lightweight dual-attention ECG network designed to capture complex ECG features essential for early HF risk prediction, despite the notable imbalance between low and high-risk groups. This network incorporates a cross-lead attention module and twelve lead-specific temporal attention modules, focusing on cross-lead interactions and each lead's local dynamics. To further alleviate model overfitting, we leverage a large language model (LLM) with a public ECG-Report dataset for pretraining on an ECG-report alignment task. The network is then fine-tuned for HF risk prediction using two specific cohorts from the UK Biobank study, focusing on patients with hypertension (UKB-HYP) and those who have had a myocardial infarction (UKB-MI).The results reveal that LLM-informed pre-training substantially enhances HF risk prediction in these cohorts. The dual-attention design not only improves interpretability but also predictive accuracy, outperforming existing competitive methods with C-index scores of 0.6349 for UKB-HYP and 0.5805 for UKB-MI. This demonstrates our method's potential in advancing HF risk assessment with clinical complex ECG data.}
\end{abstract}


\section{Introduction}
\label{SEC:introduction}
\IEEEPARstart{H}{eart} 
failure (HF) is a complex cardiovascular syndrome, where the heart fails to pump sufficient blood to meet the body's demands. Common causes of HF are cardiac structural and/or functional abnormalities, including heart attack, cardiomyopathy, and high blood pressure. HF is a chronic and progressive disease. In England, admissions due to HF have notably escalated, witnessing an increment from 65,025 in the year 2013/14 to 86,474 in 2018/19, representing a 33\% surge, as reported by the British Heart Foundation~\cite{bhf2019heartfailure}.  It has been found that around $50\%$ of deaths in HF patients presented with a sudden and unexpected pattern~\cite{tomaselli2004causes,lane_2005_prediction}, which leaves a tremendous burden on patients with HF, their families, and healthcare systems worldwide.

Preventing HF early is crucial to reduce its health and economic impacts, yet HF diagnosis often occurs late when patients have already developed serious symptoms~\cite{Yancy_2006_JACC_Clinical_presentation,Thompson_2004_JCF_Immediate}. risk~\cite{Silva_2013_P_wave,triola2005electrocardiographic,Khan_2019_JACC_10-Year,Khan_2022_Circulation,Ilkhanoff_2012_EHF}. 
\revise{A promising strategy for improving HF management is the development of risk prediction models for future HF events. These models generate a risk score for patients over a specific timeframe, taking into consideration of the patient's specific characteristics. With such a personalized risk assessment, more tailored HF management strategies and/or treatment recommendations can be provided.  In this context, the low-cost 12-lead electrocardiogram (ECG), a medical test commonly used in clinical practice, serves as a valuable resource for evaluating a patient's cardiovascular health and uncovering the risk. Recent studies have already found that several markers detected from clinically acquired 12‐lead ECG are associated with future HF events, such as prolonged QRS duration~\cite{Silva_2013_P_wave,triola2005electrocardiographic,Khan_2019_JACC_10-Year,Khan_2022_Circulation,Ilkhanoff_2012_EHF}, conduction disorders (left, right bundle-branch blocks)~\cite{Zhang_2015_Ventricular_conduction,zhang_2013_Circulation_different}. 
A significant limitation of these previous research lies in its reliance on a limited set of biomarkers (e.g., QRS duration), which are identified through predetermined rules based on ECG data. Moreover, much of this research has employed simple linear models for modeling the risk ratio associated with HF. While linear models offer a straightforward and interpretable framework for risk assessment, they may fail to capture the intricate,
complex subtleties and nuances embedded within the ECG for the early-stage risk prediction of HF.}

\revise{In recent years, deep learning-based approaches with neural networks have shown great capacities to automatically extract features from raw data and utilize them to perform a wide range of tasks. In the field of ECG analysis, deep neural networks have shown competitive performance on a wide range of tasks, including disease classification, waveform prediction, rhythm detection, mortality prediction on and automated report generation, with higher accuracy compared to traditional approaches with handcrafted features~\cite{Strodthoff_2021_Deep_Learning_for_PTB_XL,Ardeti_2023_An_overview, Hughes_2023_NPJ}. Yet, two prevalent limitations associated with deep learning-based methods are their over-fitting issues as well as poor interpretability. Overfitting occurs when a model directly memorizes the training data instead of learning  task relevant  features, reducing its ability to perform well on new, unseen data. The large number of parameters and layers in deep neural networks can further exacerbate the risk of overfitting, especially when the available training data is limited. Additionally, the complex architecture of these networks makes it difficult to trace and understand the decision-making process behind their predictions, leading to challenges in interpretability.}

In this work, we aim to develop a deep learning-based HF risk prediction model with improved feature learning, higher data efficiency, and explainability. To this end, we design a novel, lightweight ECG dual attention network. This network is capable of capturing intricate cross-lead interactions and local temporal dynamics within each lead. The dual attention mechanism also enables the visualization of lead-wise attention maps and temporal activation across each lead for improved explainability of neural network behaviors. To further alleviate the over-fitting with improved data efficiency and more importantly, to learn clinically useful representations from ECG data for higher precision, we adopt a two-stage training scheme. We first employ a large language model (LLM) to pre-train the network using a large public ECG-Report dataset covering a wide spectrum of diverse diseases and then finetune the network for the HF risk prediction task on two specific cohorts with specific risk factors collected from the UK Biobank study. Specifically, we employ ClinicalBERT~\cite{alsentzer_etal_2019_publicly}
to extract text embeddings from ECG report and force the extracted ECG features to be aligned with corresponding text features at the pretraining stage. We hypothesize that such an ECG-Text feature alignment learning paradigm can better facilitate the deep neural network to capture clinically useful patterns,  which provide a more holistic picture of patients and potentially yield more accurate risk predictions. To the best of our knowledge, this is the first work that applies LLM-informed pre-training to benefit the training of the downstream ECG-based HF risk prediction model.

To summarize, our contribution is two-fold:
\begin{enumerate}
    \item \emph{A novel deep neural network architecture to enhance both the representation learning of ECG features and the model's interpretability:} The proposed network not only yields a quantitative risk score but also offers qualitative, interpretable insights into the neural network’s reasoning through a dual attention mechanism. This unique feature acts as a transparent medium, enabling both clinicians and readers to observe and understand the intricate relationships between different ECG leads as well as the dynamic temporal patterns within each lead, highlighting the particular leads and time segments that hold the greatest importance for reliable risk prediction. 
    \item \emph{An effective model pretraining strategy with LLM:} We design an LLM-informed multi-modal pretraining task so that clinical knowledge can be transferred to the downstream risk prediction task. Training a deep risk prediction network is challenging due to a lack of sufficient event data. In this work, we advocate for the strategic use of large language models, coupled with \emph{structured ECG reports with confidence scores}, to guide the pretraining process for the benefits of more data-efficient, accurate risk prediction models.

\end{enumerate}

\section{Related work}
\subsection{Risk prediction}
Risk prediction aims to estimate the chance $h(t)$ that a patient will have a certain event in an infinitesimal time interval $D\in [t, t + \Delta t)$: $h(t)=\lim_{\Delta t \rightarrow 0} \frac{p(D \in[t, t+\Delta t) \mid D \geq t)}{\Delta t}$, given that the event has not occurred before. A common approach is the Cox proportional hazards formulation (CoxPH)~\cite{cox1972regression}. The CoxPH model is a statistical model for predicting the time-to-event outcomes based on the assumption that the hazard function $h(t)$ is proportional to a set of features or variables associated with the subject being studied. Mathematically, it can be defined as: $h(t|x) = h_0(t) \exp(g_\theta(x))$,
where $h(t|x)$ models the hazard rate at which events occur at a time $t$ taking the subject information $x$ into account, $h_0(t)$ is the baseline hazard function shared by all observations, depending only on the time $t$, and $\exp(g_\theta(x))$ is the risk score for observation $x$. Conventional CoxPH models assume that the input observation is a set of covariates (e.g., sex, age, medical history, smoking) and constrain the function $g(\cdot)$ to the linear form: $g_\theta(x)=\theta ^\intercal x = \theta_{1}\times x_1+ \theta_{2}\times x_2+\cdots+ \theta_{p}\times x_p$ with $\theta$ being the weights to the set of $p$ covariates in $x:{x_1, x_2, ..., x_p}$~\cite{cox1972regression}. 
In the existing body of literature pertaining to HF risk prediction, the predominant methodologies employ a predefined set of variables to construct a linear model, subsequently isolating variables that exhibit a high correlation for the purpose of risk prediction. This approach includes parameters such as prolonged QRS duration~\cite{Silva_2013_P_wave, triola2005electrocardiographic, Khan_2019_JACC_10-Year, Khan_2022_Circulation, Ilkhanoff_2012_EHF}, various conduction disorders (specifically left and right bundle-branch blocks)~\cite{Zhang_2015_Ventricular_conduction, zhang_2013_Circulation_different}, elongated QT intervals~\cite{triola2005electrocardiographic}, abnormalities in QRS/T angles and T wave patterns~\cite{rautaharju2013electrocardiographic, rautaharju_2006_electrocardiographic, rautaharju2007electrocardiographic, Zhang_2007_JACC_QRS, triola2005electrocardiographic}, and ST-segment depression in the V5 precordial lead~\cite{triola2005electrocardiographic, ONeal_2017_JAHA_Electrocardiographic}.

More recently, deep learning approaches such as DeepSurv~\cite{Katzman_2018_DeepSurv} consider replacing the linear function $g(\cdot)$ with a neural network $f$, which is a deep architecture  parameterized by the weights of the network $\theta$. This relaxes the risk model to a non-linear form; also, the input can be high-dimensional without needing to be linearly independent, as the neural network is capable of extracting hierarchical features for risk score estimation in a non-linear fashion. Such an approach has been found to have superior performance against traditional approaches~\cite{Katzman_2018_DeepSurv} on different risk prediction or survival prediction tasks. For example, researchers have successfully applied neural networks to automatically discover latent features from high-dimensional data for risk/survival prediction, such as whole-slide pathology images~\cite{Jiang_2023_MH,Li_2023_TMI_Survival}, 4D cardiac shape plus motion information~\cite{Bello_2019_NMI}, etc. 
In this work, we adhere to a similar philosophy and concentrate on creating neural networks designed to autonomously extract features from intricate 12-lead ECG waves, bypassing the reliance on a predefined set of ECG parameters.

\subsection{Large language model for healthcare}
Large language models (LLMs) are a class of artificial intelligence (AI) algorithms to understand human language. They can answer questions, provide summaries or translations, and create stories or poems~\cite{Singhal_2022_Nature_LLM}. Recent studies have found that LLMs can be effective in guiding representation learning on image data~\cite{Radford2021_CLIP}, enabling the knowledge transfer to several downstream vision tasks such as image classification, object detection, and segmentation. In the medical domain, such kind of multi-modal pretraining has been exploited for better understanding of imaging data, such as chest X-rays and magnetic resonance images, to benefit the downstream disease classification and medical image segmentation tasks~\cite{Liu2023_MFlag, Zhang_2023_Nature_communication, Liu2023_ICCV_CLIP, Qiu_2023_Multimodal,Turgut_2023_arxiv}. Apart from image data, most recently, concurrent works have been made on exploring the connection between natural language and signals (ECG, EEG) for better disease classification \cite{Qiu_2023_EACL,Liu_2023_ETP,han_2023_brain}. To the best of our knowledge, there is no existing risk prediction work that explores the benefit of combining ECG reports with ECG waves for better representation learning. Compared to disease classification, the task of risk prediction is more challenging due to the lack of event records for training, which amplifies the overfitting issue associated with deep neural networks. Our work provides a promising transfer learning approach to alleviate the need for a large number of event data, by utilizing LLMs and additional public large ECG-report datasets to conduct multi-modal pretraining.

\begin{figure*}[t]
\begin{center}
\includegraphics[width=\linewidth]{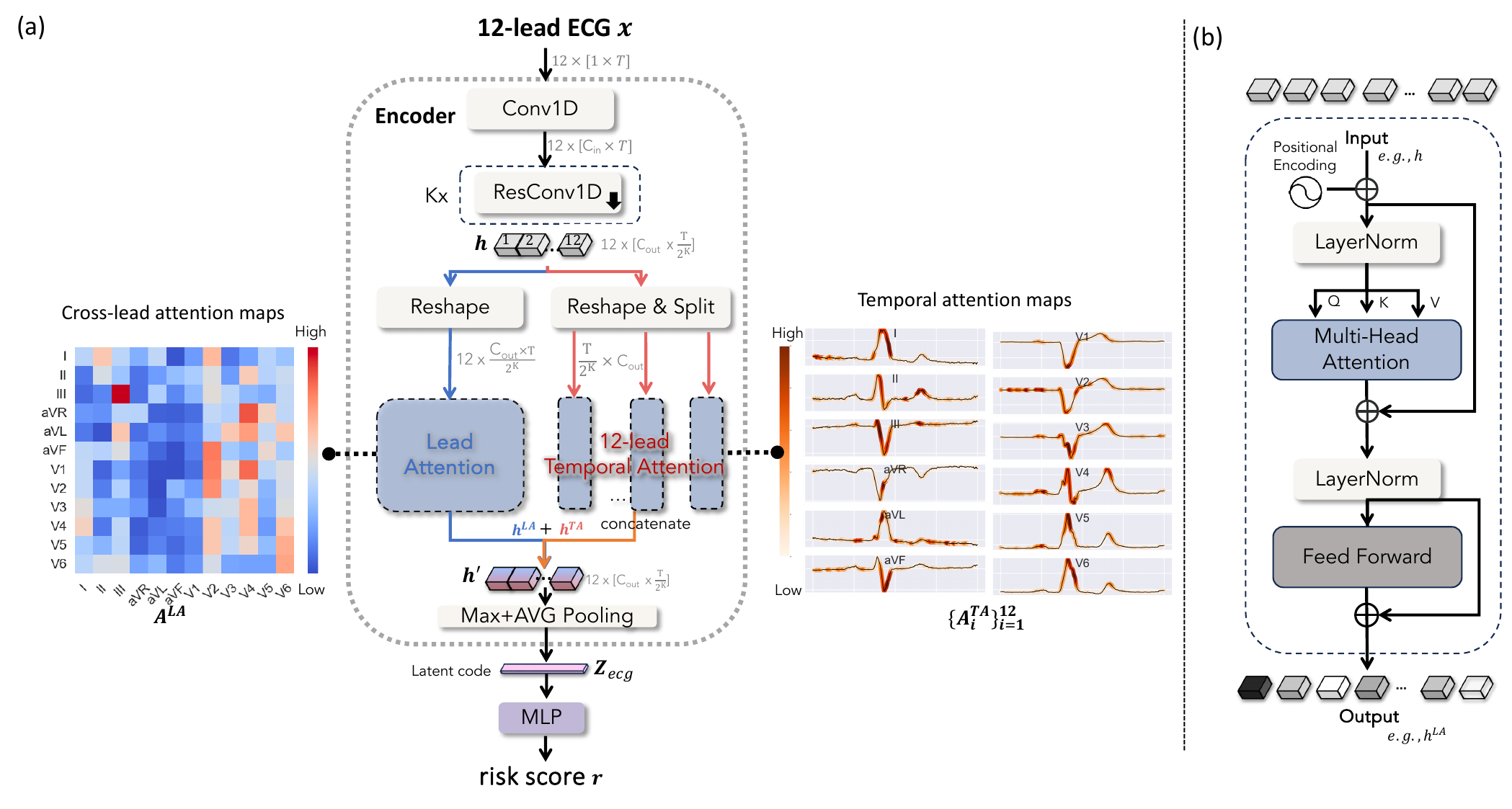}
\end{center}
\caption[ECG dual attention]{\revise{(a) Overview of the ECG dual attention encoder-based risk prediction network (ECG-DAN). A 12-lead ECG recording $\mathbf{x}$ is sent to the ECG dual attention encoder, which is capable of simultaneously extracting both cross-lead relationships as well as temporal dynamic patterns within each lead, for better feature aggregation. Then, features from two routes are added and then sent to a max-average pooling layer, producing a flattened feature vector $\mathbf{z}_{ecg}$. Finally, we employ a multi-layer perceptron (MLP) module to map from a high-dimensional feature space into a risk score (scalar) $r$ for heart failure. (b) Overview of the core attention module used in lead attention and temporal attention modules. See Sec.~\ref{SEC:methodology}B for more details.}}
\label{fig:ecg_attention}
\vspace{-4pt}
\end{figure*}

\section{Methods}
\label{SEC:methodology}
\subsection{Overview}
Assume we have a dataset of $N$ triplets $\{\mathbf{x}_i, \delta_i, t_i \}_{i=1}^{N}$ to record the HF events in a  population. Here, $\mathbf{x}_i\in \Reals ^{12\times T}$ is a 12-lead ECG signal (I, II, III, AVL, AVR, AVF, V1-6)  with a recording length of $T$; $\delta_i$ indicates whether there is known date of HF; $t_i$ is the number of month to the censoring time if there is no reported HF event during the follow‐up period ($\delta_i=0$, right censored) or the number of months until the patient was diagnosed with HF during the follow-up ($\delta_i=1$, uncensored). The objective is to have a risk prediction model $f$ parameterized by $\theta$ so that it can predict a patient’s risk of HF given the patient's ECG data. To this end, we design an ECG dual attention network (ECG-DAN), as shown in Fig.~\ref{fig:ecg_attention}(a), where the input is a 12-lead ECG median wave recording $\mathbf{x}$ and the output of the network is a single node $r$, which estimates the risk score $r= f(\mathbf{x};\theta) $. 
\subsection{ECG-DAN network}
A signature of ECG-DAN is a dual attention module,  which is designed to extract morphological and spatial changes and relationships across different leads as well as the temporal dynamics inside each lead for a more comprehensive understanding of the heart's electrical activity. We first process each lead signal via a group of 1D convolution layers for noise filtering and feature extraction, which gives $C_{in}$-channel features for each lead at each time point. We then employ a set of $K$ residual blocks with $K$ $2\times$ down-sampling layers along the temporal dimension to  extract features at different scales.  The output for each lead consists of feature maps with $C_{out}$ channels and a reduced time dimension of $\frac{T}{2^K}$: $\mathbf{h}_i\in \Reals^{C_{out}\times \frac{T}{2^K}}$. Those features from all 12 leads $\mathbf{h}:\{\mathbf{h}_1, \mathbf{h}_2,...,\mathbf{h}_{12}\}$ are then chained and sent to the Lead Attention (LA) block, facilitating the learning of cross-lead interactions to enhance feature aggregation globally. Concurrently, a 12-lead Temporal Attention (TA) component is employed to capture crucial temporal patterns within each lead $\mathbf{h}_i$ along the time dimension. Here, for each lead $\mathbf{h}_i$, we apply an individual temporal attention module {TA}$_{i}$ across the time domain separately, given the fact that different leads correspond to different directions of cardiac activation within three-dimensional space. The outcomes of the two attention modules are added and then summed with a concat-pooling module, producing concatenated flattened features from max-pooling and average pooling operations following common practice from previous ECG analysis work~\cite{Strodthoff_2021_Deep_Learning_for_PTB_XL}: 
\begin{equation}
\begin{aligned}
\small
\mathbf{h}' &= \mathbf{h}^{LA}+\mathbf{h}^{TA}\\
 \mathbf{z}_{ecg} &= \textrm{Concat}(
\textrm{Maxpooling}(\mathbf{h}');
\textrm{Avg-pooling} (\mathbf{h}')).
\end{aligned}
\end{equation}
With this latent feature $\mathbf{z}_{ecg}$, we then send it to a multi-layer perceptron network (MLP), which consists of three linear layers, reducing the feature by half and then projecting the feature into a 3-dimensional feature space, finally, regressing it to a risk score $r$.

Next, we will explain the two key modules for improved feature learning and model explainability: the lead attention module and the temporal attention, in more detail.
\subsubsection{Lead attention}
 We apply the multi-head attention-based encoder module introduced in transformers~\cite{Vaswani_2017_NIPS} to capture the cross-lead interactions. Specifically, we first reshape the feature matrix  ($\mathbf{h} \in \Reals^{12\times \frac{C_{out}T}{2^K}})$ (12 is number of leads) where each lead feature is treated as a token. As shown in Fig.~\ref{fig:ecg_attention}(b), a sinusoid positional encoding $\operatorname{PE}$~\cite{Vaswani_2017_NIPS} is added to the input feature $\mathbf{h} \leftarrow \text{LayerNorm}(\mathbf{h} + \operatorname{PE}(\mathbf{h}))$ to encode the contextual information with layer normalization~\cite{Ba2016_LayerNorm}, followed by a multi-head self-attention module to capture cross-lead interactions. At a high level, the input to the attention are: key matrix ($\mathbf{K}\in \Reals^{12\times \frac{C_{out}T}{2^K}}$), query matrix  ($\mathbf{Q}\in \Reals^{12\times \frac{C_{out}T}{2^K}}$), and value matrix ($\mathbf{V}\in \Reals^{12\times \frac{C_{out}T}{2^K}}$) where key and query matrices are used to compute the weights to re-weight the value matrix. Mathematically, it can be expressed as:
 \begin{equation}
 \small
 \mathbf{h}^{LA}=\operatorname{softmax}\left(\frac{\mathbf{Q K}^{\top}}{\sqrt{D_k}}\right) \mathbf{V}=\mathbf{A}^{LA}\mathbf{V}
\end{equation}
where $\mathbf{A}^{LA} \in \Reals^{12 \times 12}$ is the attention weight matrix and $D_k$ is the feature dimension ($\frac{C_{out}T}{2^K}$). We use the same input $\mathbf{h}$ for computing $\mathbf{K},\mathbf{Q},\mathbf{V}$~\footnote{In practice, for computational efficiency, following \cite{Vaswani_2017_NIPS}, we apply the multi-head attention trick, which first linearly projects the queries, keys, and values $h$ times with different learnable 
linear projection matrices to lower dimensions ($D_k/h$) respectively to compute the attention matrix and re-weight the projected value matrix, then concatenate all the output $h$ heads to recover the feature dimension.}. 

In addition to the self-attention, a fully connected feed-forward network (FFN) is applied to refine the re-weighted features with residual connection:
\begin{equation}
\begin{aligned} 
\small
\mathbf{h}^{LA} & =\text{LayerNorm}(\mathbf{h}+\mathbf{h}^{LA}) \\
\mathbf{h}^{LA}& = \operatorname{FFN}(\mathbf{h}^{LA}) = \mathbf{h}^{LA}+\max (0, (\mathbf{h}^{LA} W_1+b_1) W_2+b_2)
\end{aligned}
\end{equation}
 where $W_1, W_2$ and $b_1, b_2$ are weights and biases of two linear layers in FFN; $\text{LayerNorm}$ is a layer normalization layer \cite{Ba2016_LayerNorm} for feature normalization. In this way, the output feature is adjusted with all the information from other leads into account.

\subsubsection{12-lead temporal attention (TA)}
The structure of the temporal attention module is very similar to the lead attention. The only difference is that now we have $12$ separate attention modules for different lead features so that each module is adapted to a specific lead to extract temporal dynamic features locally. Specifically, we split the feature $\textbf{h}$ along the lead dimension, and for each lead feature $\textbf{h}_i$, we first add sinusoid positional encoding $PE$ information along the time domain and then process each lead with a separate temporal attention module where each time point is treated as a token. In other words, the input becomes a sequence of $C_{out}$-dimensional features with a sequence length of $l=\frac{T}{2^K}$ and the corresponding temporal attention matrix for lead $i$ becomes: $\mathbf{A}^{TA}_{i} \in \Reals^{l \times l}$.  \revise{Similar to the lead attention module process, we reweight and normalize the lead feature $\mathbf{h}_i$ with the generated temporal attention matrix and then employ an FFN module for feature refinement for each lead. The output of the temporal attention is a concatenation of output from the 12-lead temporal attention modules. The whole process can be defined as follows:
\begin{equation}
\begin{aligned} 
\mathbf{h}_i & \leftarrow \text{LayerNorm}(\mathbf{h}_i + \operatorname{PE}(\mathbf{h}_i))\\
\mathbf{h}_i^{TA} & =\text{LayerNorm}(\mathbf{h}_i+\mathbf{A}^{TA}_{i}\mathbf{h}_i^{\top}) \\
\mathbf{h}_i^{TA} &= \operatorname{FFN}_i(\mathbf{h}_i^{TA})\\
\mathbf{h}^{TA} &= \operatorname{Concat}({\mathbf{h}_1^{TA},\mathbf{h}_2^{TA},..., \mathbf{h}_{12}^{TA}})^\top.
\end{aligned}
\end{equation}
}
\subsection{Training}
\begin{figure*}[t]
\begin{center}
\includegraphics[width=0.9\textwidth]{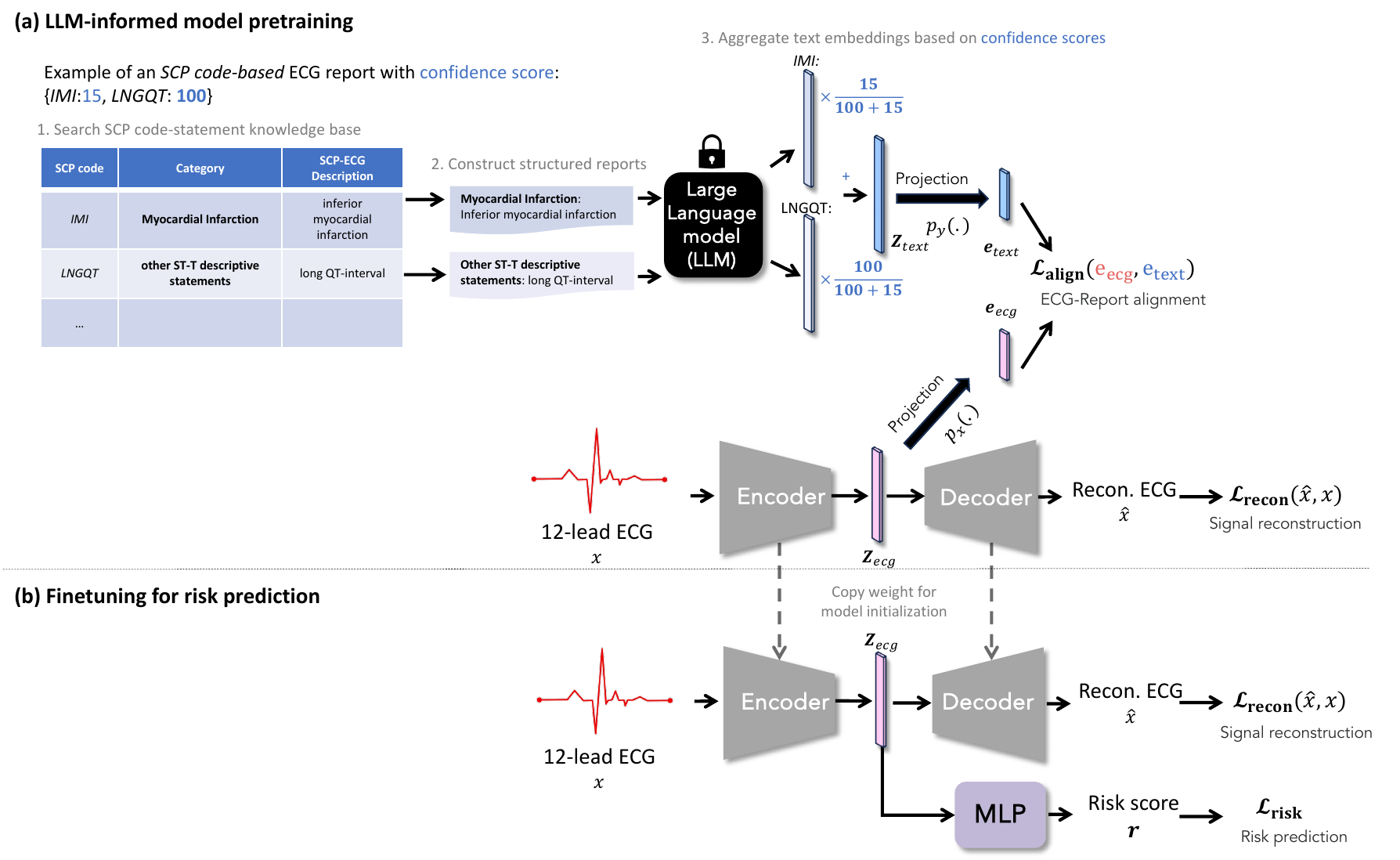}
\end{center}
\caption[Language model informed pretraining]{\revise{Training overview. Our model is a) first pretrained on the ECG-Report alignment task and the signal reconstruction task on a large-scale public dataset (PTB-XL~\cite{Wagner2020_PTB_XL,Strodthoff_2023_PTBXL}), and  then b) finetuned on the heart failure risk prediction task with two specific cohorts from the UK Biobank where the future HF event data is available. Here, in PTB-XL dataset, each report has been abstracted to a set of SCP codes with SCP-ECG statement description and confidence score (annotated by human experts). We construct a structured report based on SCP-ECG protocol~\cite{Wagner2020_PTB_XL} and then send it to a \emph{frozen} LLM to extract clinical knowledge for better representation learning guidance. As one ECG may have multiple SCP-code relavant statements, we extract text features separately and then use confidence-based reweighting to aggregate features for feature summation.} See below texts for more details.}
\label{fig:pretraining}
\vspace{-3mm}
\end{figure*}
Deep learning-based risk prediction often struggles with small datasets, particularly when events are rare, as in our case where HF events are below $5\%$. To overcome this and prevent overfitting, we adopt a two-stage training approach, see Fig.~\ref{fig:pretraining}. Initially, we train our network on a large, diverse public ECG dataset.  Of note, this dataset does not contain any HF records for risk prediction. After that, we initialize the model with the pretrained parameters and then conduct fine-tuning, focusing on risk prediction for particular populations with documented HF from the UK Biobank study~\cite{Sudlow_2015_UKB}. Subsequently, we fine-tune the pretrained model on the UK Biobank data, specifically targeting HF risk prediction. The pretraining incorporates human-verified ECG reports to align features with clinical knowledge, aiming to improve the model's ability to discern pathological ECG patterns relevant to HF risk.

\subsubsection{Large language model informed model pre-training}
\label{SEC:llm-pretrain}
For pretraining, we employed the PTB-XL dataset~\cite{Wagner2020_PTB_XL, Strodthoff_2023_PTBXL}, a large, expert-verified collection of $21,799$ clinical 12-lead ECGs with accompanying text reports. It features annotations by cardiologists according to the SCP-ECG standard and classifies waveforms into five categories: Normal (NORM), Myocardial Infarction (MI), ST/T Change (STTC), Conduction Disturbance (CD), and Hypertrophy, with possible overlap due to concurrent conditions. We followed the dataset creators' protocol, using folds 1-8 for training and folds 9-10 for validation and testing during pretraining~\cite{Wagner2020_PTB_XL}.
\\
\textbf{Extracting latent text code $\mathbf{z}_{text}$ from reports using large language model:}
We employ an LLM to extract the knowledge embedded in ECG reports. Specifically, we use the medical domain language model: BioClinical BERT~\cite{alsentzer-etal-2019-publicly}~\footnote{\url{https://huggingface.co/emilyalsentzer/Bio_ClinicalBERT}}, which has been trained on a large number of electronic health records from MIMIC III \cite{Johnson_2016_MIMIC}. We consider two ways of extracting text embeddings $\mathbf{z}_{text}$:
\begin{itemize}
    \item \textbf{Latent text code from raw ECG report (raw)}: For a piece of ECG report $y$ (in English)~\footnote{Since the original reports are written in a mixture of German and English, we used the open-source machine translation tool: EasyNMT~\footnote{\url{https://github.com/UKPLab/EasyNMT}} for batch translation following~\cite{Qiu_2023_EACL} and then used ChatGPT\footnote{\url{https://chat.openai.com/}} to further refine those failed cases with a prompt: \textit{translate the ECG report into English:$\#text$}.}, we simply feed it to the LLM to get $\mathbf{z}_{text}=\operatorname{LLM}(y)$, following~\cite{Liu_2023_ETP,Qiu_2023_Multimodal}.   
    \item \textbf{Latent text code from structured ECG report weighted by confidence (structured with confidence)}: The original PTB-XL dataset also provides ECG-SCP codes generated from ECG reports. Specifically, each report has been abstracted to a set of SCP codes with SCP-ECG statement description and confidence score (annotated by human experts). 
    In this case, we build a structured sentence with linked SCP statement category information and SCP description for each SCP code: $\mathbf{y}(SCP)$ = `\emph{$\{\#$Statement Category(SCP)\}:$\{\#$SCP-ECG Statement Description(SCP)}\}'\footnote{Database containing SCP code-statement mappings can be found at~\url{https://physionet.org/content/ptb-xl/1.0.1/scp_statements.csv} and~\url{https://physionet.org/content/ptb-xl/1.0.1/ptbxl_database.csv}}. \revise{For example, as shown in Fig~\ref{fig:pretraining}(a), given an SCP code: LNGQT, the input becomes: `other ST-T descriptive statements: long QT-interval'. We send this type of structured input to the LLM, which gives an embedding $\mathbf{z^{SCP}} = \operatorname{LLM}(\mathbf{y}({SCP}))$ for each SCP code. It is important to note that a single recording may encompass multiple SCP codes. In that case, multiple SCP embeddings are initially derived. These embeddings are then aggregated with weights based on corresponding confidence scores $c$ to get the text embedding for the corresponding ECG: $\mathbf{z}_{text} = \sum_j \frac{ c_j \mathbf{z}^{SCP}_{j}}{c^{sum}}$, where $c^{sum}$ is the sum of all confidence scores for all corresponding statements for the input ECG signal.}
\end{itemize}
\subsubsection{ECG-report alignment loss}
To align the ECG to report, similar to~\cite{Radford2021_CLIP}, we first project the latent ECG code $\mathbf{z}_{ecg}$ and the latent text code $\mathbf{z}_{text}$  to $\mathbf{e}_{ecg}$, $\mathbf{e}_{text}$ with two learnable projection functions ${p}_x$, ${p}_y$ respectively, so that the two embeddings $\mathbf{e}_{ecg}={p}_x(\mathbf{z}_{ecg})$, $\mathbf{e}_{text} = {p}_y(\mathbf{z}_{text})$ are of the same dimension, as shown in Fig.~\ref{fig:pretraining} (a). We use a distance loss $\mathcal{D}$ to quantify the dissimilarity between the two: 
\begin{equation}
    \Loss_{align} = \frac{1}{n}\sum_{i=1}^{n}[{\mathcal{D}((\mathbf{e}_{ecg}, \mathbf{e}_{text})_i)}],
\end{equation}
where $\mathcal{D}$ is a cosine embedding metric function: $\mathcal{D}(\mathbf{e}_{ecg}, \mathbf{e}_{text})=1-\operatorname{cos}(\mathbf{e}_{ecg}, \mathbf{e}_{text})$, a metric that is commonly used for measuring the distance between two embeddings. The loss value is an average value over a  batch of $N$ paired embeddings.
\subsubsection{Pretraining loss}
The total loss for pretraining is a combination of the ECG-report alignment loss and a signal reconstruction loss, defined as:
\begin{equation}
\begin{split}
\small
\Loss_{pretraining} =  \Loss_{recon} +\Loss_{align}\\
 \Loss_{recon} =  \frac{1}{n}\sum_{i=1}^{n}(\|\mathbf{x}_i - \hat{\mathbf{x}}_i\|^2)
 \end{split}
\end{equation}
where we compute the mean-squared-error between input $\mathbf{x_i}$ and reconstructed signals $\hat{\mathbf{x}}_i$ for every input signal $\mathbf{x_i}$ in a batch and average them. Adding a signal reconstruction loss is necessary as it can help uncover latent generic features in the ECG signal and serve as a regularization term to avoid latent space collapse problems. During pre-training, we only update the parameters in the ECG encoder, ECG decoder, and two projectors, while keeping the parameters of the language model \emph{frozen} for training stability and efficiency, as suggested by prior work~\cite{Liu2023_MFlag}. Detailed network structures can be found in the Appendix. 

\subsubsection{Finetuning loss}
After pre-training, we copy the model weights to initialize the risk prediction model, see Fig.~\ref{fig:pretraining}(b) and then finetune the risk prediction network. The finetuning loss is also a multi-task loss, including the self-supervised signal reconstruction loss, and a risk loss~\cite{Katzman_2018_DeepSurv}, which aims to minimize the average negative log partial likelihood of the set of uncensored patients ($\delta_i=1$: developed into HF during the follow-up). 
The risk loss is defined as:
\begin{equation}
\small
   \Loss_{risk} = - \frac{1}{n_{\delta=1}}\sum_{i:\delta_i=1}\left [f_\theta(\boldsymbol{x}_i) - \log \sum_{j: t_j \geq t_i}\exp(f_\theta(\boldsymbol{x}_j))\right]
\end{equation}
where ${n_{\delta=1}}$ is the number of uncensored subjects in a batch, and $f_\theta(\boldsymbol{x})$ is a predicted risk score.

The total loss is then defined as: 
\begin{equation}
\small
   \Loss_{finetuning} =\alpha \Loss_{recon} + (1-\alpha) \Loss_{risk}
\label{loss:finetune}
\end{equation}
where $\alpha$ is a trade-off parameter to balance the contribution of two losses. 

During model optimization, an issue with  $\Loss_{risk}$  is that it can be very sensitive to the number of uncensored subjects ${n_{\delta=1}}$ (number of subjects who develop HF in the follow-up) in the training batch. This causes training instability problems if the number becomes zero or jumps from large to small and vice versa between batches.  In real-world datasets, this issue is amplified due to a high-class imbalance between censored and uncensored subjects. To address this problem, we, therefore, perform a modified version of stochastic gradient descent for model optimization, where 
we select a batch of $n$ observations with stratified random sampling to ensure every batch maintains a comparable ratio of censored to uncensored observations, mirroring the ratio found in the entire training population.

\section{Experiments}
\label{SEC:Experiments}
\subsection{Study population}
\label{SEC:Study population}
In this study, we focus on two populations: patients with hypertension (HYP) and MI, which are highly related to disease progression to HF. Subjects are selected from the UK Biobank dataset (UKB), which is a large-scale biomedical database containing genetic, demographic, and disease information and is regularly updated with comprehensive follow-up studies, from approx. 500,000 subjects. The UKB dataset consists of a large portion of healthy subjects as well as those with a range of cardiovascular and other diseases. To assess the future risk of HF, our analysis is confined to individuals who have  electrocardiogram (ECG) together with imaging data available and have not been diagnosed with HF either prior to or at the time of the ECG evaluation.

\subsubsection{UKB-HYP} HF-free subjects with prevalent HYP (had HYP before or during the ECG examination) at baseline time from the UKB dataset are studied~\cite{Sudlow_2015_UKB}. We identified 11,581 HF-free HYP subjects. Follow‐up time was defined as the time from the baseline ECG measurement until a diagnosis of HF or death or the end of follow‐up (January 5, 2023). Most ECG recordings together with images were taken between 2014 and 2021. Records with less than two years of follow-up time were excluded. Among 11581 participants, 162 ($1.2\%$) participants developed HF. The median follow-up time is 56 months (4.7 years), and the maximum follow-up time is 87 months (7.3 years).

\subsubsection{UKB-MI} HF-free subjects at baseline with prevalent MI records are studied. Similar to the above selection procedure, we identified 800 subjects.  Among them, 32 subjects ($4\%$) developed HF during the follow-up period. The median follow-up time is 53 months (4.4 years), and the maximum follow-up time is 83 months (6.9 years). 
\begin{table}[!h]
\centering
\caption{Statistics of studied population(s).}
\label{tab:dataset}
\resizebox{\columnwidth}{!}{%
\begin{tabular}{@{}c@{ }@{ }c@{ }@{ }c@{ }@{ }c@{ }@{ }c@{}}
\toprule
& \multirow{2}{*}{characteristics} & \multirow{2}{*}{total} & \multicolumn{2}{c}{had HF during follow-up?} \\ \cmidrule(l){4-5} 
&                                 &                        & Yes                      & No                \\ \midrule
UKB-HYP & Age at examination$^\#$ & 66.31 (45-82)           & 69.57 (49-81)            & 66.27 (45-82)     \\
& Sex (men/women)                 & 6730/4851              & 109/53 (1.6\%/1.1\%)     & 6621/4798         \\ \midrule
UKB-MI & Age at examination$^\#$ & 68.21 (49-81)          & 67.28 (50-79)            & 68.25 (49-81)     \\
& Sex  (men/women)                 & 650/150                & 27/5 (4.3\%/3.4\%)       & 623/145           \\ \bottomrule
\multicolumn{5}{@{}l}{$^\#$Values represent mean (minimum-maximum).}
\end{tabular}%
}
\end{table}

Code lists used for the retrieval of disease information can be found in the Appendix.

\subsection{Implementation Details}
\label{SEC:Implementation details}
For ECG signals, we used 12-lead median waveforms (covering a single beat)  with a frequency of 500~Hz as input. Each lead is preprocessed with $z$-score normalization and then zero-padded to a length of $1024$. We employ 5 residual blocks ($K=5$, $C_{in}=4$, $C_{out}=16$) to obtain multi-scale features. The dimension of latent ECG feature $\mathbf{z}_{ecg}$ is 512, and the dimension of projected features $\mathbf{e}_{ecg}$, $\mathbf{e}_{text}$ is 128. The default convolutional kernel size is 5. Further network details are provided in the Appendix. We used a batch size of 128 ($n=128$) for model updates at pre-training, with random lead masking for data augmentation~\cite{Nejedly_2022_CINC_winner}. AdamW optimizer~\cite{Loshchilov_2019_AdamW} with stratified batches was used for training. For fine-tuning in post-MI risk prediction (UKB-MI), we maintained a batch size of 128. Given the low HF incidence ($<2\%$) in UKB-HYP, we increased the batch size to 1024 to include enough uncensored subjects for calculating $\mathcal{L}_{risk}$. \revise{Batch-wise dropout was used to stabilize fine-tuning~\cite{Jiang_2023_MH}as well as avoid overfitting}. The loss function's weighting parameter ($\alpha$) is set at 0.5. Pre-training and fine-tuning were conducted over 300 and 100 epochs, respectively, to ensure convergence.
\subsection{Evaluation metrics and evaluation method}
We report the concordance index (C-index)~\cite{Harrell1982_C_index} as the primary evaluation metric. This metric measures the accuracy of the ranking of predicted time based on the radio of concordant pairs:$
\text { C-index }=\frac{\text { \# concordant pairs }}{\text{ \# concordant pairs }+ \text{ \# discordant pairs }}
$. A concordant pair refers to when the predicted time ranking of two subjects aligns with their actual ranking, while a discordant pair means the opposite. A pair $(i, j)$ is concordant if $t_i < t_j$ and risk scores $r_i> r_j$, or vice versa. A value of 1 denotes perfect prediction, while a value of 0.5 indicates a prediction quality equivalent to random guessing. 

\revise{\textbf{Robust evaluation with multiple two-fold stratified cross-valiation.} Due to the scarcity of uncensored subjects (with HF events), using conventional deep learning data splits (e.g, five-fold cross validation or 7:1:2 for training, validation and testing) would result in too few HF events for reliable C-index evaluation. Hence, we opted for a 1:1 training-to-testing ratio with \emph{stratified} cross-validation, aligned with patient HF status during follow-up. This approach constituted a two-fold \emph{stratified} cross-validation, which also helps to maintain the same proportion of each class as in the original dataset in each split. To further avoid over-fitting, $20\%$ of the training data was randomly allocated as a validation set for hyper-parameter search (e.g. choice of the learning rate) and model selection. Specifically, we split the dataset into two folds, one for training and validation, the rest for testing. This process is then repeated once more, with the roles of the two parts reversed. To ensure that the result is not biased by the splitting, we perform the above procedure five times, each with a different split across two datasets (UKB-HYP and UKB-MI). The final model performance is reported as the average (with standard deviation) of these 10 trials.}


\section{Results}
\label{SEC:Results}
\subsection{LLM-informed pre-training improves the accuracy of the downstream risk prediction}
\begin{table}[t]
\caption{Comparison of risk prediction performance using different pre-training tasks. All experiments were performed using the same proposed network architecture. Reported values are average C-index over 5 cross-validations using different random splits.}
\centering
\label{tab:comparison_on_different_pretraining_tasks}
\resizebox{0.5\textwidth}{!}{%
\begin{tabular}{@{}c@{ }@{ }c@{}@{}c@{}}
\toprule
\multirow{3}{*}{ Pretraining Tasks } & \multicolumn{2}{c}{Study population} \\
& Hypertension & Myocardial Infarction \\
& (UKB-HYP) & (UKB-MI) \\\midrule
- & 0.6122 (0.0190) & 0.5065 (0.0776) \\ \midrule
SR & 0.6327 (0.0165) & 0.5069 (0.0770) \\
SR + Classification & 0.6370 (0.0216) & 0.5220 (0.0475) \\\midrule
SR + ECG-R Alignment (raw) & 0.6088 (0.0189) & 0.5796 (0.0570) \\
SR + ECG-R Alignment & \multirow{2}{*}{\textbf{0.6349 (0.0156)}} & \multirow{2}{*}{\textbf{0.5805 (0.0580)}} \\
(structured with confidence) &  &  \\\bottomrule
\multicolumn{3}{@{}l}{SR: Signal Reconstruction; Classification: ECG Disease Classification;}\\
\multicolumn{3}{@{}l}{ECG-R: ECG-Report.}
\end{tabular}
}
\end{table}

We first compared our proposed pretraining strategy against various pretraining tasks: 
\begin{itemize}
    \item No pretraining (represented by a hyphen), serving as a baseline, with networks trained for 400 epochs (equivalent to the sum of pretraining and finetuning epochs);
\item Pretraining on signal reconstruction only;
\item Pretraining on signal reconstruction combined with multi-label disease classification (NORM, MI, STTC, CD, hypertrophy), using a cross-entropy loss for classification;
\item Pretraining on signal reconstruction and ECG-Report alignment using raw text reports;
\item Our proposed method, pretraining on signal reconstruction and ECG-Report alignment with structured report and confidence information, as detailed in Sec.~\ref{SEC:llm-pretrain}.
\end{itemize}
Results are shown in Table~\ref{tab:comparison_on_different_pretraining_tasks}. 
It is clear that the models pretrained using the proposed strategy  (signal reconstruction + ECG2Text alignment (structured w/ confidence)) consistently obtain the highest accuracy among the two tasks, obtaining the average C-index: 0.6349 (0.0156) on UKB-HYP and 0.5805 (0.0580) on UKB-MI.

\subsection{Comparison study on risk prediction using different network architectures}
\begin{table*}[!ht]
\caption{\revise{Comparison of risk prediction performance using different types of deep neural networks. By default, all network has been initialized with pre-trained on the proposed ECG-Report alignment and reconstruction tasks (300 epochs) and then finetuned on the risk prediction task (100 epochs). For fair comparison, models without any pre-training were trained for 400 epochs (300+100). To avoid overfitting, models with the highest C-index score on the validation set were selected as the final model. Reported values are average C-index over 5 cross-validations using different random splits (10 trials in total).}}
\label{tab:comparison_on_network}
\centering
\resizebox{0.8\textwidth}{!}{%
\color{black}\begin{tabular}{@{}cclcc@{}}
\toprule
\multirow{2}{*}{Network architectures} &
\multirow{2}{*}{$\#$ parameters} &
\multicolumn{1}{c}{\multirow{2}{*}{LLM-informed Pretraining}} &
\multicolumn{2}{c}{Study population} \\ \cmidrule(l){4-5} 
                                   &                & \multicolumn{1}{c}{} & Hypertension (UKB-HYP) & Myocardial Infarction (UKB-MI) \\ \midrule

CNN-VAE~(\cite{Yuling_2023,Marcel_2022_ISBI,Marcel_2022_Frontiers,Lei_2022_STACOM})                            & 7.0M           & \xmark                    & 0.6215 (0.0237)       & 0.5675 (0.0353)                \\
CNN-VAE~(\cite{Yuling_2023,Marcel_2022_ISBI,Marcel_2022_Frontiers,Lei_2022_STACOM})                            & 7.0M           & \yesmark                  & 0.6346 (0.0218)       & 0.5484 (0.0269)               \\
XResNet1D (~\cite{Strodthoff_2021_Deep_Learning_for_PTB_XL})                 & 1.9M           & \xmark                         & 0.5601 (0.0273)        & 0.5357 (0.0485)                \\
XResNet1D~(\cite{Strodthoff_2021_Deep_Learning_for_PTB_XL})            & 1.9M           & \yesmark                        & 0.6234 (0.0143)        & 0.5431 (0.0387)                \\
ECG dual attention (The proposed) & \textbf{1.4M } & \xmark                    & 0.6122 (0.0190)        & 0.5065 (0.0776)                \\
ECG dual attention (The proposed) &
  \textbf{1.4M} &
  \yesmark    &
  \textbf{0.6349 (0.0156)} &
  \textbf{0.5805 (0.0580)} \\ \bottomrule
\end{tabular}
}
\end{table*}
We further compare model performance using different encoder architectures, including 
1) the encoder used in a variational autoencoder (VAE)-like network architecture, which has been found effective for ECG signal feature representation learning in previous works~\cite{Yuling_2023,Marcel_2022_ISBI,Marcel_2022_Frontiers,Lei_2022_STACOM};
2) XResNet1D~\cite{Strodthoff_2021_Deep_Learning_for_PTB_XL}, which is the top performing network architecture on a wide range of different ECG analysis tasks such as ECG disease classification, age regression, and form/rhythm prediction on the public PTB-XL benchmark dataset~\cite{Wagner2020_PTB_XL} and ICBEB2018 dataset~\cite{Liu_2018_ICBEB2018}. We report their performance trained w/o and w/ the proposed language model informed pre-training strategy with structured SCP report in Table~\ref{tab:comparison_on_network}. The proposed ECG attention network has the fewest parameters. Yet, models with this type of network architecture and the proposed pre-training strategy obtain the highest C-index scores on both two tasks. 

\subsection{Comparison study between traditional ECG parameter-based risk prediction vs ECG dual attention}
We further compare our method to the traditional method with well-established ECG parameters. Specifically, we collect a set of ECG parameters from the ECG signals, which have been identified in previous studies for incident HF prediction and mortality estimation~\cite{ Silva_2013_P_wave,Khan_2019_JACC_10-Year,Khan_2022_Circulation, Zhang_2015_Ventricular_conduction,zhang_2013_Circulation_different,rautaharju2013electrocardiographic,rautaharju_2006_electrocardiographic,rautaharju2007electrocardiographic,Zhang_2007_JACC_QRS,triola2005electrocardiographic, ONeal_2017_JAHA_Electrocardiographic}. All of these parameters were automatically extracted by the ECG device (using the GE CardioSoft V6~\footnote{\url{https://biobank.ndph.ox.ac.uk/ukb/ukb/docs/CardiosoftFormatECG.pdf}}) with supporting evidence from previous work: 1) Ventricular rate;
2) Left-axis deviation; 3) Right-axis deviation; 4) Prolonged P-wave duration ($>120~\textit{ms}$)~\cite{ONeal_2017_JAHA_Electrocardiographic}; 5) Prolonged PR interval ($>200~\textit{ms}$)~\cite{ONeal_2017_JAHA_Electrocardiographic}; 6) Prolonged QRS duration ($>100~\textit{ms}$)~\cite{ONeal_2017_JAHA_Electrocardiographic}; 7) Prolonged QT interval ($\geq460~\text{(women)}/450~\text{(men)}~\textit{ms}$ using the Framingham formula)~\cite{Sagie_1992_Framingham}; 8) Delayed intrinsicoid deflection (DID time) (the maximum value in leads V5 and V6 $>50~\textit{ms}$)~\cite{ONeal_2016_Electrocardiographic_time_to_ID}; 9) Abnormal P-wave axis (values outside the range of $0^{\circ}$ and $75^{\circ}$)~\cite{Li_2014_P_wave}; 10) Left ventricular hypertrophy~\cite{Devereux_1984_JACC_Cornell}; 11) Abnormal QRS-T angle ($>77^{\circ}~\text{(women)}/88^{\circ}~\text{(men)}$)~\cite{ONeal_2017_JAHA_Electrocardiographic};
12) Low QRS voltage  \cite{Prineas_2009_The_Minnesota_Code};
13) ST-T abnormality \cite{Prineas_2009_The_Minnesota_Code};
14) Right bundle‐branch block~\cite{Prineas_2009_The_Minnesota_Code};
15) Left bundle‐branch block~\cite{Prineas_2009_The_Minnesota_Code}.
\begin{table}[!h]
\caption{Comparison of risk prediction performance using traditional ECG parameter-based and our deep learning model (ECG dual attention). Reported values are average C-index over five times of 2-fold cross-validations using different random splits (10 trials in total).}
\label{tab:comparison_to_traditional_method}
\centering
\resizebox{\columnwidth}{!}{%
\color{black}\begin{tabular}{@{}ccccc@{}}
\toprule
\multirow{2}{*}{Method} &
  \multirow{2}{*}{Input} &
  \multirow{2}{*}{\begin{tabular}[c]{@{}c@{}}Model \\ Type\end{tabular}} &
  \multicolumn{2}{c}{Study population} \\ \cmidrule(l){4-5} 
 &
   &
   &
  \begin{tabular}[c]{@{}c@{}}Hypertension  \\ (UKB-HYP)\end{tabular} &
  \begin{tabular}[c]{@{}c@{}}Myocardial Infarction \\ (UKB-MI)\end{tabular} \\ \midrule
Tradiational &
  \begin{tabular}[c]{@{}c@{}}15 \\ biomarkers\end{tabular} &
  Linear &
  0.6149 (0.0125) &
  0.5398 (0.0200) \\
\begin{tabular}[c]{@{}c@{}}ECG dual attention\\  (the proposed)\end{tabular} &
  \begin{tabular}[c]{@{}c@{}}12-lead\\ Median waves\end{tabular} &
  Nonlinear &
  \textbf{0.6349 (0.0156)} &
  \textbf{0.5805 (0.0580)} \\ \bottomrule
\end{tabular}
}
\end{table}

\begin{figure}[t]
\begin{center}
\includegraphics[width=\columnwidth]{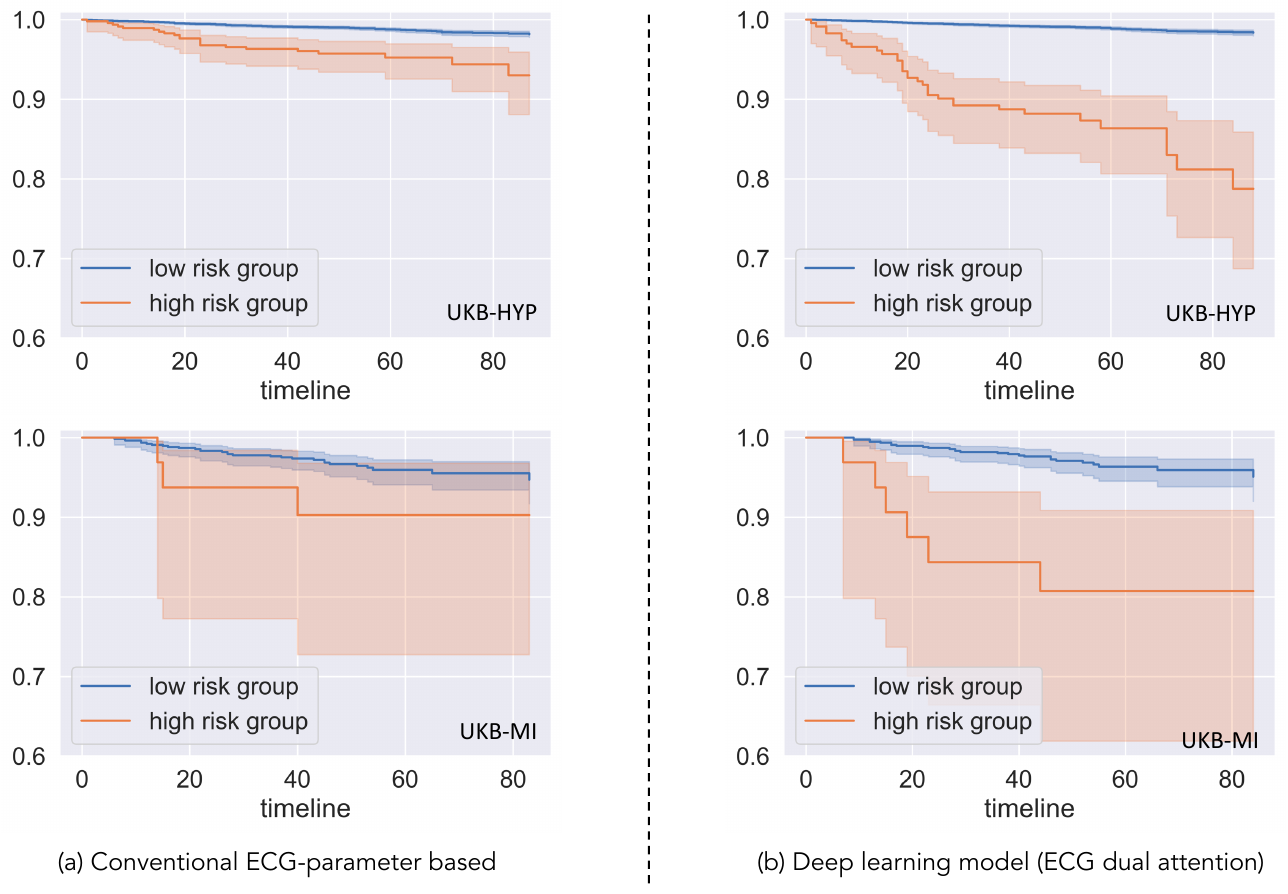}
\end{center}
\caption[risk curve visualization]{Kaplan-Meier risk curve plots for a) conventional parameter model using a composite of ECG measurements, and b) the proposed deep learning-based risk prediction model with ECG dual attention blocks where the model has been pretrained on the ECG-report alignment task. For both models,
patients were divided into low- and high-risk groups with a cutoff value referenced from the top 98$^{th}$ percentile (for UKB-HYP) or top 96$^{th}$ percentile (UKB-MI) risk scores predicted by the model, reflecting the statistics of the datasets in Table~\ref{tab:dataset}.}
\label{fig:risk_curve_visualization}
\vspace{-4mm}
\end{figure}
\begin{figure*}[tb]
\begin{center}
\includegraphics[width=0.8\textwidth]{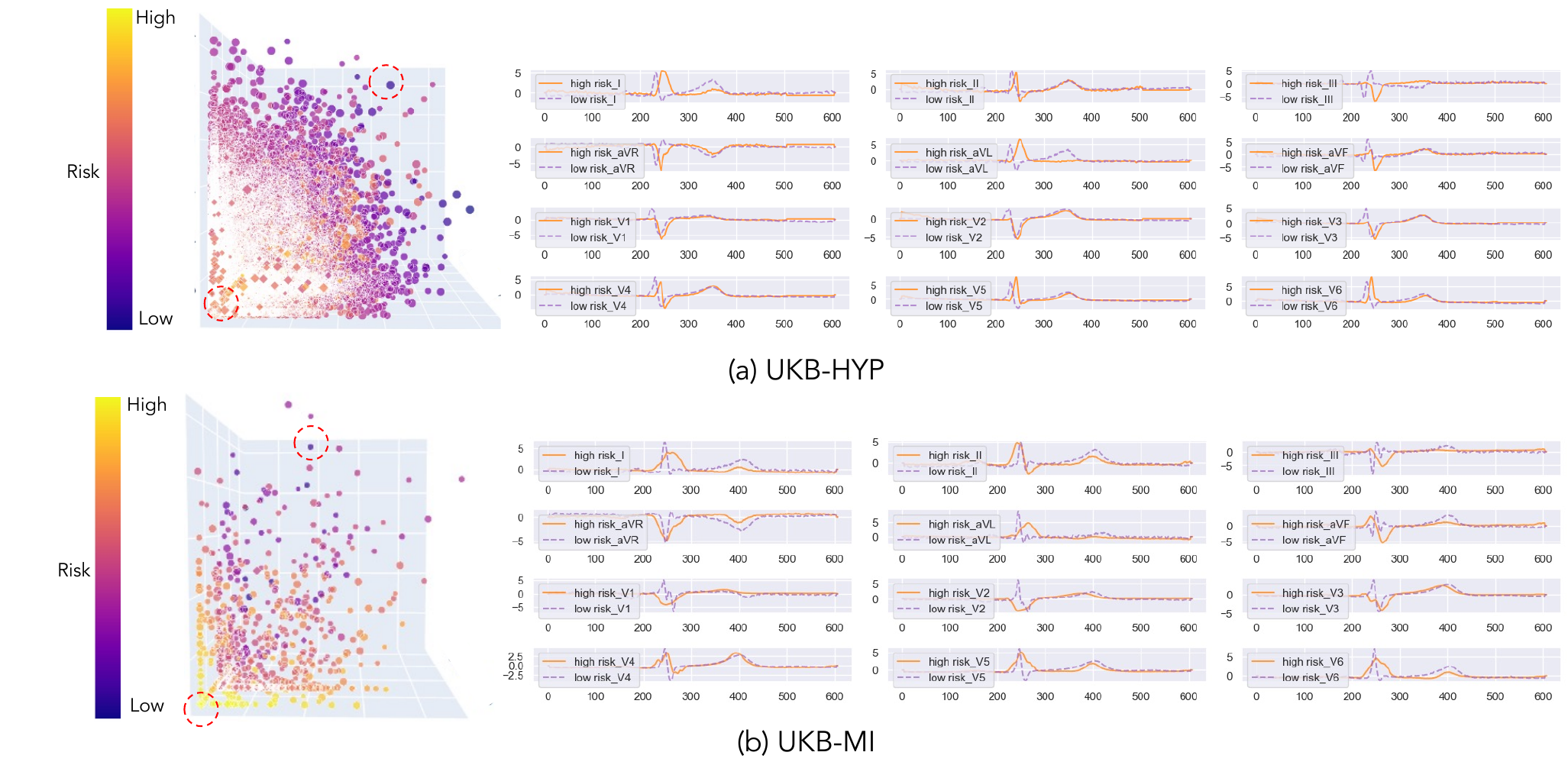}
\end{center}
\caption[risk code visualization]{3D visualization of the last 3-dim hidden feature  learned in the risk prediction subnetwork along with the visualization of input ECG waves with lowest predicted risk score (dark purple) and highest predicted risk score (light yellow) on the a) UKB-HYP and b) UKB-MI datasets.}
\label{fig:risk_code_visualization}
\vspace{-2mm}
\end{figure*}

Following~\cite{ONeal_2017_JAHA_Electrocardiographic}, we fit these ECG parameters into the conventional CoxPH linear regression model~\cite{cox1972regression} and use it as a strong competing model for HF risk prediction. Table~\ref{tab:comparison_to_traditional_method} shows the performances using our approach and the traditional ECG parameter-based approach, and Fig.~\ref{fig:risk_curve_visualization} shows Kaplan-Meier plots which depict the survival probability estimates over time, stratified by risk groups defined by each model’s predictions. We further plot the features learned in the last hidden layer ($\in \Reals^3$) of ECGs from the risk prediction branch in the proposed network and representative ECG waves with the lowest/highest risk score in Fig.~\ref{fig:risk_code_visualization}.

\subsection{Dual attention mechanism improves model stability and interpretability}
\subsubsection{Quantitative ablation study}
\begin{table}[!h]
\centering
\caption{Ablation Experiments. Numbers in bold represent the highest values, while numbers with underlines denote the second highest.}
\label{tab:ablation_on_attention}
\resizebox{\columnwidth}{!}{%
\begin{tabular}{@{}lcc@{}}
\toprule
\multirow{2}{*}{\textbf{Method}} & \multicolumn{2}{c}{\textbf{Study population}}                                         \\
 & \begin{tabular}[c]{@{}c@{}}Hypertension  \\ (UKB-HYP)\end{tabular} & \begin{tabular}[c]{@{}c@{}}Myocardial Infarction  \\ (UKB-MI)\end{tabular} \\ \midrule
w/o time attention module      & 0.6038 (0.0243)          & \textbf{0.5855 (0.0353)} \\
w/o lead attention module      & 0.6018 (0.0418)          & 0.4920 (0.0884)          \\
w/o lead+time attention module & {\ul 0.6167 (0.0105)}    & 0.5195 (0.0539)          \\ \midrule
The proposed                   & \textbf{0.6349 (0.0156)} & {\ul 0.5805 (0.0580)}    \\ \bottomrule
\end{tabular}%
}
\end{table}
We also evaluate the impact of the dual attention modules on whether they can help to enhance the accuracy of risk prediction. We conduct the ablation study experiments using the same training strategy with the same network but either the lead and/or the temporal attention module removed. Results are shown in Table~\ref{tab:ablation_on_attention}. It can be observed that the propose model containing both attention modules consistently produces the most stable performance, with the best performance on the UKB-HYP and the second best performance on the UKB-MI, yielding the best average performance across the two datasets. 

\subsubsection{Lead attention matrix visualization}
\begin{figure}[!ht]
\begin{center}
\includegraphics[width=\columnwidth]{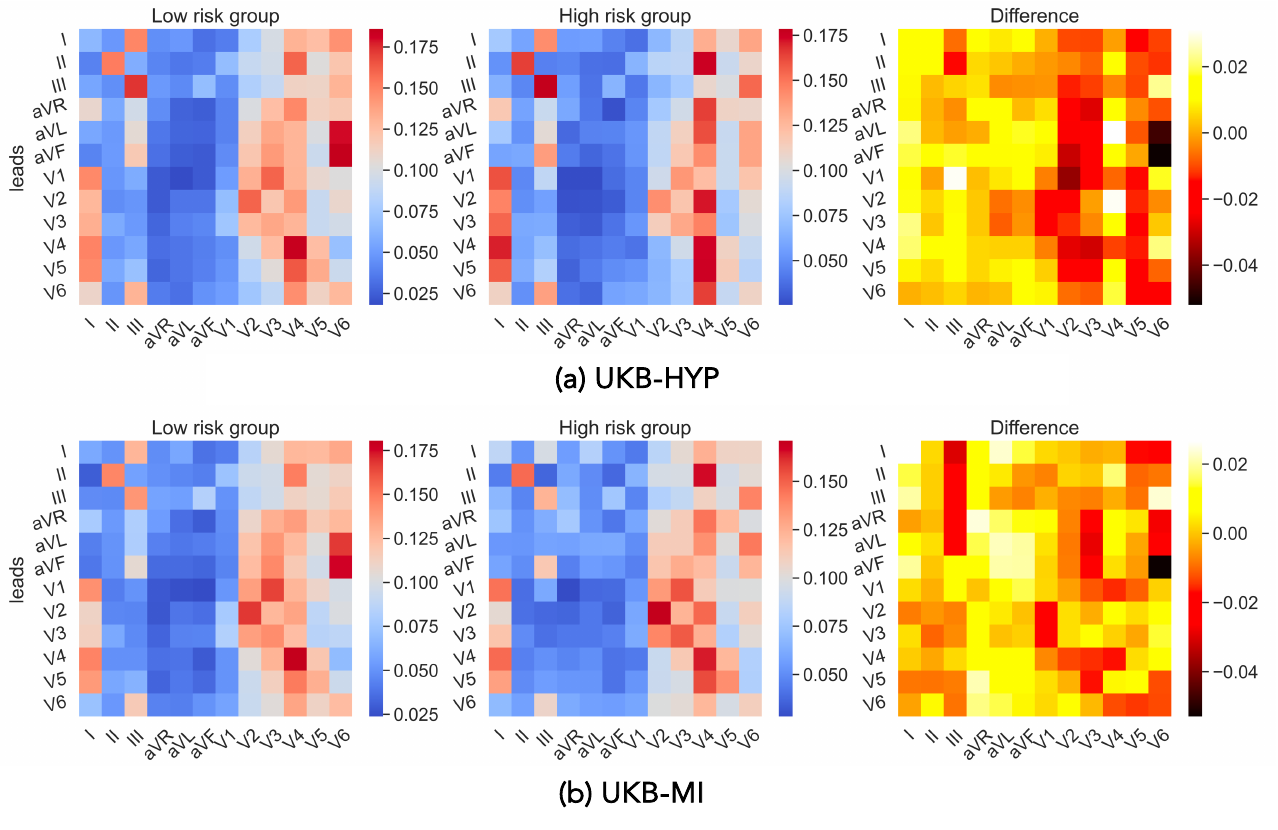}
\end{center}
\caption[LeadAttentionVisualization]{\revise{Visualization of lead attention patterns and differences between low-risk and high-risk groups across two, different populations: UKB-HYP and UKB-MI.}}
\label{fig:lead_attention_visualization}
\end{figure}

We further visualize the average lead attention matrix at the population level, averaging matrices in different risk groups. For both datasets, patients are divided into low- and high-risk groups with a cutoff value
referenced from the top 98$^{th}$ percentile (for UKB-HYP) or top 96$^{th}$ percentile (UKB-MI) risk scores predicted by the model, where the threshold is chosen based on the statistics of the datasets. Of note, all risk scores and attention matrices are obtained using the same cross-validation strategy, where the predicted risk is computed by averaging predictions from models trained with data excluding that subject. Figure~\ref{fig:lead_attention_visualization} illustrates the contribution of each lead for feature re-weighting across all 12-lead features for different populations. We can find similar patterns though the underlying study cohorts are different. 

\subsubsection{Temporal attention activation map visualization}
\begin{figure*}[t]
\begin{center}
\includegraphics[width=0.67\textwidth]{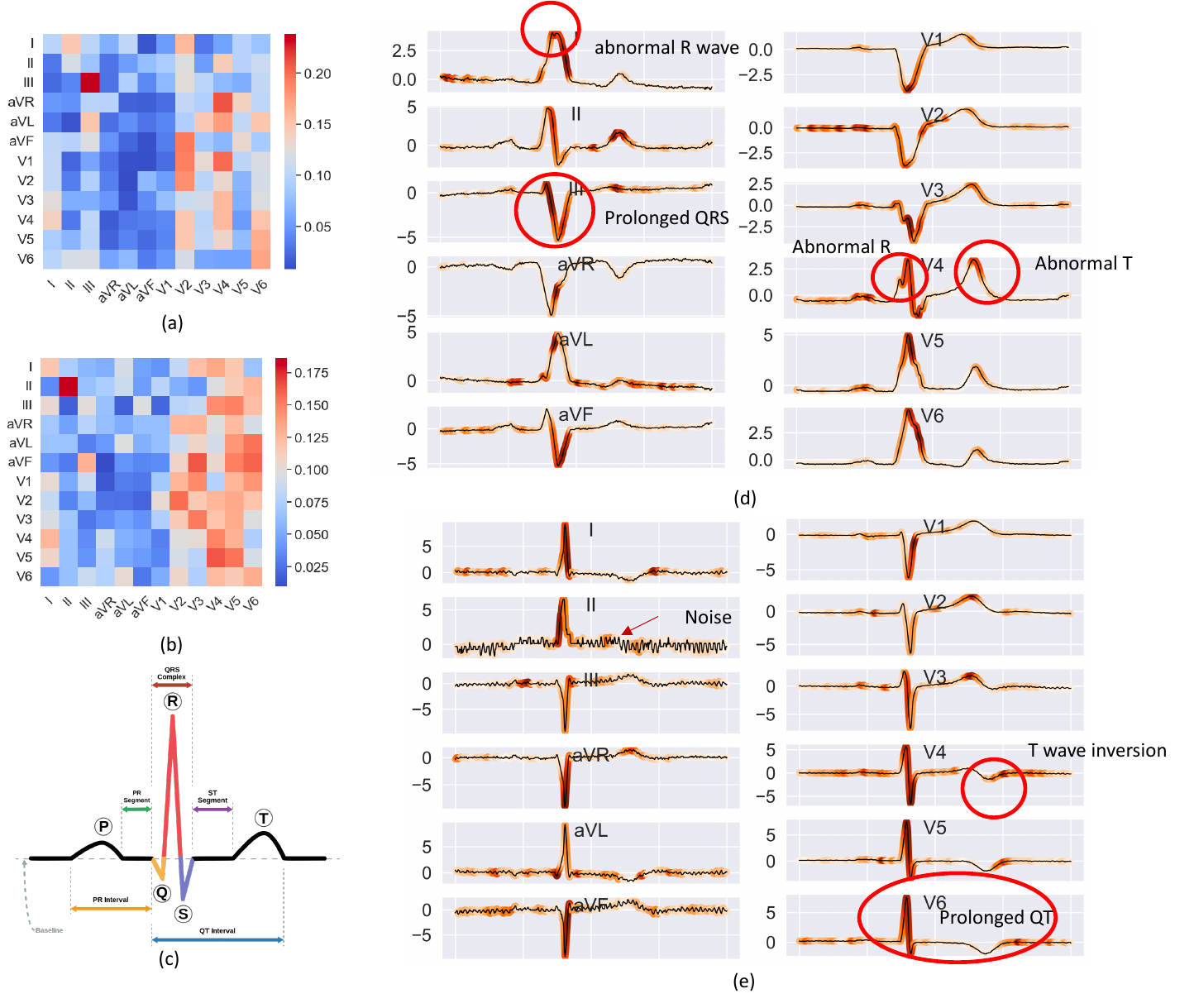}
\end{center}
\caption[TimeAttentionVisualization]{\revise{Visualization of cross-lead (a,b) and 12-lead temporal attention maps (d,e) obtained from a high HF risk ECG with HYP (a,d) and a high HF risk ECGs with MI (b,e). (c) is a schematic standard ECG for illustrative purpose. Source: {Wikimedia Commons}}.}
\label{fig:time_attention_visualization}
\end{figure*}
To visualize the temporal attention matrix $\mathbf{A}_i^{TA}$ in the original ECG time length $T$, we adapt GradCAM for ECG leads to highlight the model's focus. We condense the attention matrix by summing column values for attention scores across time, and then map them to the ECG input features, creating Grad-CAM maps. These maps, weighted by their attention scores, reveal the network's focus areas on the ECG. More details of implementation can be found in the Appendix. The visualization is presented in Fig.~\ref{fig:time_attention_visualization}.


\section{Discussion}
\label{SEC:Discussion}
\begin{figure}[t]
\begin{center}
\includegraphics[width=0.6\columnwidth]{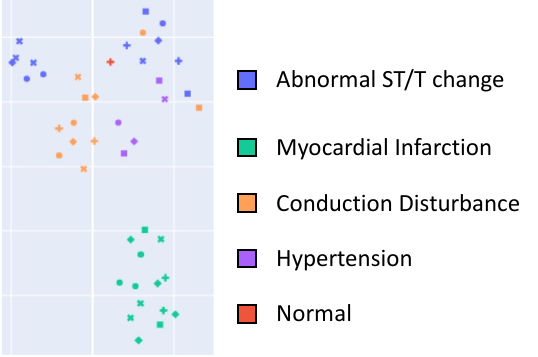}
\end{center}
\caption[SCP code visualization]{U-map visualization of latent code embeddings $\mathbf{z}^{SCP}$ from the large language model using different structured SCP statements. Different colors represent the categorization of statements with disease labels.}
\label{fig:SCP_code_visualization}
\vspace{-2pt}
\end{figure}
\textbf{Importance of language informed pretraining:}
In the experiments, we first studied the impact of different pretraining tasks for downstreaming risk predictions and highlighted the value of LLM-informed pretraining for risk prediction in Table~\ref{tab:comparison_on_different_pretraining_tasks}. In general, pre-training tasks enhance the risk prediction performance compared to those without pretraining, especially on the smaller dataset (UKB-MI). It is interesting to see that the performance of the deep risk prediction model (average C-index: 0.5065) can be inferior to the traditional approach (average C-index 0.5398) if without proper pretraining, see Table~\ref{tab:comparison_on_different_pretraining_tasks} and Table~\ref{tab:comparison_to_traditional_method}. This indicates that pre-training is important to alleviate model over-fitting. Our study further highlights the critical role of targeted model pretraining in identifying both generic and pathological features for downstream HF risk prediction. Pretraining on structured ECG reports outperforms methods using simple disease labels or unstructured text inputs. We credit this improved performance to the integration of detailed context from structured reports and supplementary confidence information. Figure~\ref{fig:SCP_code_visualization} shows the UMAP~\cite{McInnes2018_UMAP} visualization of latent text codes of structured reports $\mathbf{z^{SCP}}$ extracted by LLM, which suggests that LLM distinguishes between disease-specific embeddings (e.g., separated clusters) and captures the interrelations among various diseases. For instance, the embedding for the normal or healthy category (red dots) is nearer to ST/T wave change embeddings (blue dots) and notably distant from those for MI (green dots), which is aligned with the fact that ST/T variations can be non-pathological, unlike the distinct disease pattern of MI.

\textbf{ECG dual attention network exhibits high effectiveness of risk stratification:}
By comparing the results in Table~\ref{tab:comparison_on_network}, it is clear that the suggested approach surpasses leading deep learning approaches in terms of both computational costs and accuracy. \revise{Interestingly, we observe that language-informed pretraining, while continuously boosting the performance on the ResNet-based structure and our attention model, does not enhance the CNN-VAE-based model's performance. This phenomenon probably attributed to the over-parameterization in the CNN-VAE networks (7.0M). In that case, the prioritization of strong regularization to shape the  latent representation to like a Gaussian distribution to ensure feature independence~\cite{Kingma_2013_VAE} becomes crucial. Such regularization might conflict with the language-informed training process.}

\revise{Results in Table~\ref{tab:comparison_to_traditional_method} also show that our approach surpasses the performance of a traditional methodology, which relies on a predefined set of ECG parameters with a simple linear model, as well as current leading deep learning-based models in terms of both computational costs and performance. The high overlapping region between high-risk and low-risk curves of MI populations (left bottom) when using the traditional approach in Fig.~\ref{fig:risk_curve_visualization} reveals the challenge when the underlying population is limited. By contrast, our method can consistently stratify the patients into different risks with a clear gap.} Figure~\ref{fig:risk_code_visualization} demonstrates that our network effectively discriminates between low-risk (upper right, purple dots) and high-risk ECGs (bottom left, yellow dots) in the latent space, correlating high-risk ECGs with clinical markers like prolonged QRS duration and longer QT intervals, in alignment with established studies~\cite{triola2005electrocardiographic,Silva_2013_P_wave,Khan_2019_JACC_10-Year,Khan_2022_Circulation,Ilkhanoff_2012_EHF}.

\textbf{ECG dual attention network exhibits high interpret-ability:} Furthermore, the proposed two dual attention modules offer model interpretability, a feature highly valued in the clinical setting. The integration of a dual attention mechanism serves as a window into the network's thought process, allowing clinicians and readers alike to visualize and comprehend the complex interactions between different ECG leads, as shown in Fig.~\ref{fig:lead_attention_visualization}. First, we found the three augmented leads (aVR, aVL, aVF) contributed the least (see blue regions). We believe that this may be because aVR, aVL, and aVF can be derived from Lead I, II, and III using Goldberger’s equations~\cite{ecgwaves2023}, thus containing redundant information for feature aggregation and risk prediction. Second, we found that the network typically pays more attention to chest lead I and precordial leads  (V2-V6) (see red regions). In clinical practice, comparing the morphological changes across the precordial leads (V1-V6) can help to identify ST/T wave abnormality, which could be an indicator of future HF~\cite{rautaharju2013electrocardiographic,rautaharju_2006_electrocardiographic,rautaharju2007electrocardiographic,Zhang_2007_JACC_QRS, triola2005electrocardiographic} and sudden cardiac death~\cite{Tikkanen2015-yt}. On the other hand, upon comparing the attention patterns derived from high-risk groups to those from low-risk groups, as illustrated in the column titled `Difference` in Fig.~\ref{fig:lead_attention_visualization}, we observed elevated activation values within the high-risk groups. This observation underscores the network's heightened sensitivity in identifying abnormal features.

\revise{For better understanding, we visualizes the cross-lead and temporal activation maps of two high-risk cases in Figure~\ref{fig:time_attention_visualization}. For temporal ones, we apply GradCAM~\cite{Ramprasaath_GRADCAM} to mapping the low-resolution temporal  attention maps back to the original input signal level (see appendix for more details). It can be observed that cross-lead module provides an overview to identify the uniqueness or exploit synergy or cross-lead interactions among various leads, whereas temporal lead attention focuses on identifying local areas in each lead important to the prediction.  For example, Leads III and II stand out in the cross-lead attention maps (a) and (b) respectively, highlighting unique pathological pattern (prolonged negative QRS in lead III, see (d) or strong abnormal noises in lead II, see original signals in (e), in contrast to other leads. Additionally, the orange diagonal clusters in panel (b) showcases good R wave progression from V2 to V6, see (e). On the other hand, the temporal attention maps in (d) and (e) reveal more intricate details, showing the neural network's capability to focus on clinically significant features such as P, Q, S, T waves, and R peaks, even with strong noise presented, and highlight pathological abnormalities like abnormal R peaks, T wave iregularities and prolonged QRS and QT intervals. This implies that the dual attention network can autonomously discover clinically relevant biomarkers from ECG data without explicit being taught during its training phase. Future research will aim at collaborating with medical professionals to verify potential novel biomarkers using these visual tools.}

\textbf{Limitations:} One of the limitations of the current work is that it only considers information from ECG signals. In the future, we will consider extending our approach to a broader spectrum of data. The input features can be blood test results, demographic (age, sex, ethnicity) \cite{Yuling_2023}, smoking, chronic disease condition such as diabetes, imaging-derived features~\cite{Lei_2022_STACOM,Marcel_2022_Frontiers,Grafton-Clarke2023_ESC_2023} as well as genetic information to create a more holistic and accurate characterization of the patient~\cite{Khan_2019_JACC_10-Year,Khan_2022_Circulation,Cikes_2019_EJHF_Machine,Biton_2021_Atrial}. Moreover, it is interesting to increase the inclusivity and diversity of our study by considering a broader population base.

\textbf{Broader Impact:}\revise{We believe the proposed LLM-informed pretraining scheme is not limited to the risk prediction task. It also holds the potential to inspire new approaches and applications for ECG signal analysis, such as disease diagnosis and ECG segmentation. The LLM model, utilizing textual reports, acts as an additional source of contextual supervision, which guides the ECG network to better understand and extract complex patterns. We would also like to highlight that the utilized public pre-training dataset which comprises a diverse and comprehensive pathologies is crucial for model generalization. In the future, it is also interesting to further enhance the model interpretability through the incorporation of the generative capabilities of LLM, providing structured report for further explanation}.
\section{Conclusion}
\label{SEC:Conclusion}

This paper presents a study with a novel ECG dual attention network for ECG-based HF risk prediction. This network distinguishes itself through its unique blend of being both lightweight and efficient, outperforming existing models in the field. Its standout feature is the ability to generate not just a quantitative risk score but also to provide a qualitative interpretation of its internal processes. This is achieved through the generation of attention visualization maps, which span across and delve within individual leads, offering a granular view of the network focus and decision-making process. We hope to contribute to the development of more transparent and interpretable AI-assisted systems, fostering trust and facilitating broader adoption in clinical settings. Additionally, the study presents a multi-modal pretraining approach for risk prediction models, which leverages external public ECG reports and confidence scores from \emph{diverse} populations, combined with advanced large language models, to address the challenges posed by limited future event data in risk prediction tasks.

\section*{Acknowledgment}
This research has been conducted using the UK Biobank Resource under Application Number `40161'. 
The authors express no conflict of interest.
\printbibliography

@book{patent,
  author    = {Hauthor Surname1}, 
  title     = {The title of the work},
  publisher = {Patent number},
  address   = {Patent country},
  year      = 2010,
}

@string{MICCAI      = {Medical Image Computing and Computer Assisted Intervention}}

@string{STACOM       = {International Workshop on Statistical Atlases and Computational Models of the Heart}}

@string{ISBI = {International Symposium on Biomedical Imaging}}

@string{Arxiv ={arXiv Preprint}}

@string{NIPS       = {Conference on Neural Information Processing Systems}}

@string{ICLR        = {International Conference on Learning Representations}}

@string{ICCV = {International Conference on Computer Vision}}

@string{ICLR = {International Conference on Learning Representations}}

@string{NMI        = {Nature Machine Intelligence
}}

@BOOK{Prineas_2009_The_Minnesota_Code,
  title     = "The Minnesota Code Manual of Electrocardiographic Findings",
  author    = "Prineas, Ronald J and Crow, Richard S and Zhang, Zhu-Ming",
  publisher = "Springer London",
  month     =  nov,
  year      =  2009,
  language  = "en"
}

@ARTICLE{ONeal_2016_Electrocardiographic_time_to_ID,
  title    = "Electrocardiographic Time to Intrinsicoid Deflection and Heart
              Failure: The {Multi-Ethnic} Study of Atherosclerosis",
  author   = "O'Neal, Wesley T and Qureshi, Waqas T and Nazarian, Saman and
              Kawel-Boehm, Nadine and Bluemke, David A and Lima, Joao A C and
              Soliman, Elsayed Z",
  journal  = "Clinical Cardiology",
  volume   =  39,
  number   =  9,
  pages    = "531--536",
  month    =  sep,
  year     =  2016,
  language = "en"
}

@ARTICLE{Sagie_1992_Framingham,
  title    = "An improved method for adjusting the {QT} interval for heart rate
              (the {Framingham Heart Study})",
  author   = "Sagie, A and Larson, M G and Goldberg, R J and Bengtson, J R and
              Levy, D",
  journal  = "The American Journal of Cardiology",
  volume   =  70,
  number   =  7,
  pages    = "797--801",
  month    =  sep,
  year     =  1992,
  language = "en"
}

@inproceedings{Liu2023_MFlag,
   title         = "{M-FLAG}: Medical {Vision-Language} Pre-training with Frozen
                   Language Models and Latent Space Geometry Optimization",
  author        = "Liu, Che and Cheng, Sibo and Chen, Chen and Qiao, Mengyun
                   and Zhang, Weitong and Shah, Anand and Bai, Wenjia and
                   Arcucci, Rossella",
  month         =  jul,
  year          =  2023,
  booktitle = MICCAI,
  eprint        = "2307.08347"
}

@inproceedings{alsentzer-etal-2019-publicly,
    title = "Publicly Available Clinical {BERT} Embeddings",
    author = "Alsentzer, Emily  and
      Murphy, John  and
      Boag, William  and
      Weng, Wei-Hung  and
      Jindi, Di  and
      Naumann, Tristan  and
      McDermott, Matthew",
    booktitle = "Proceedings of the 2nd Clinical Natural Language Processing Workshop",
    month = jun,
    year = "2019",
    address = "Minneapolis, Minnesota, USA",
    publisher = "Association for Computational Linguistics",
    url = "https://aclanthology.org/W19-1909",
    doi = "10.18653/v1/W19-1909",
    pages = "72--78",
}

@ARTICLE{Johnson_2016_MIMIC,
  title    = "{MIMIC-III}, a freely accessible critical care database",
  author   = "Johnson, Alistair E W and Pollard, Tom J and Shen, Lu and Lehman,
              Li-Wei H and Feng, Mengling and Ghassemi, Mohammad and Moody,
              Benjamin and Szolovits, Peter and Celi, Leo Anthony and Mark,
              Roger G",
  journal  = "Scientific Data",
  volume   =  3,
  pages    = "160035",
  month    =  may,
  year     =  2016,
  language = "en"
}

@ARTICLE{Li_2014_P_wave,
  title    = "Effect of electrocardiographic {P}-wave axis on mortality",
  author   = "Li, Yabing and Shah, Amit J and Soliman, Elsayed Z",
  journal  = "The American Journal of Cardiology",
  volume   =  113,
  number   =  2,
  pages    = "372--376",
  month    =  jan,
  year     =  2014,
  language = "en"
}

@ARTICLE{Grafton-Clarke2023_ESC_2023,
  title    = "Cardiac magnetic resonance left ventricular filling pressure is
              linked to symptoms, signs and prognosis in heart failure",
  author   = "Grafton-Clarke, Ciaran and Garg, Pankaj and Swift, Andrew J and
              Alabed, Samer and Thomson, Ross and Aung, Nay and Chambers,
              Bradley and Klassen, Joel and Levelt, Eylem and Farley, Jonathan
              and Greenwood, John P and Plein, Sven and Swoboda, Peter P",
  journal  = "ESC heart failure",
  volume   =  10,
  number   =  5,
  pages    = "3067--3076",
  month    =  oct,
  year     =  2023,
  language = "en"
}

@ARTICLE{Hughes_2023_NPJ,
  title    = "A deep learning-based electrocardiogram risk score for long term
              cardiovascular death and disease",
  author   = "Hughes, J Weston and Tooley, James and Torres Soto, Jessica and
              Ostropolets, Anna and Poterucha, Tim and Christensen, Matthew Kai
              and Yuan, Neal and Ehlert, Ben and Kaur, Dhamanpreet and Kang,
              Guson and Rogers, Albert and Narayan, Sanjiv and Elias, Pierre
              and Ouyang, David and Ashley, Euan and Zou, James and Perez,
              Marco V",
  journal  = "NPJ digital medicine",
  volume   =  6,
  number   =  1,
  pages    = "169",
  month    =  sep,
  year     =  2023,
  language = "en"
}

@ARTICLE{Devereux_1984_JACC_Cornell,
  title    = "Electrocardiographic detection of left ventricular hypertrophy
              using echocardiographic determination of left ventricular mass as
              the reference standard. Comparison of standard criteria, computer
              diagnosis and physician interpretation",
  author   = "Devereux, R B and Casale, P N and Eisenberg, R R and Miller, D H
              and Kligfield, P",
  journal  = "Journal of the American College of Cardiology",
  volume   =  3,
  number   =  1,
  pages    = "82--87",
  month    =  jan,
  year     =  1984,
  language = "en"
}

@ARTICLE{Ardeti_2023_An_overview,
  title   = "An overview on state-of-the-art electrocardiogram signal
             processing methods: Traditional to {AI-based} approaches",
  author  = "Ardeti, Venkata Anuhya and Kolluru, Venkata Ratnam and Varghese,
             George Tom and Patjoshi, Rajesh Kumar",
  journal = "Expert Systems with Applications",
  volume  =  217,
  pages   = "119561",
  month   =  may,
  year    =  2023
}

@ARTICLE{ONeal_2017_JAHA_Electrocardiographic,
  title    = "Electrocardiographic Predictors of Heart Failure With Reduced
              Versus Preserved Ejection Fraction: The {Multi-Ethnic} Study of
              Atherosclerosis",
  author   = "O'Neal, Wesley T and Mazur, Matylda and Bertoni, Alain G and
              Bluemke, David A and Al-Mallah, Mouaz H and Lima, Joao A C and
              Kitzman, Dalane and Soliman, Elsayed Z",
  journal  = "Journal of the American Heart Association",
  volume   =  6,
  number   =  6,
  month    =  may,
  year     =  2017,
  language = "en"
}

@inproceedings{alsentzer_etal_2019_publicly,
    title = "Publicly Available Clinical {BERT} Embeddings",
    author = "Alsentzer, Emily  and
      Murphy, John  and
      Boag, William  and
      Weng, Wei-Hung  and
      Jin, Di  and
      Naumann, Tristan  and
      McDermott, Matthew",
    booktitle = "Proceedings of the 2nd Clinical Natural Language Processing Workshop",
    month = jun,
    year = "2019",
    address = "Minneapolis, Minnesota, USA",
    publisher = "Association for Computational Linguistics",
    url = "https://www.aclweb.org/anthology/W19-1909",
    doi = "10.18653/v1/W19-1909",
    pages = "72--78"
}

@ARTICLE{Tikkanen2015-yt,
  title    = "Electrocardiographic {T} Wave Abnormalities and the Risk of
              Sudden Cardiac Death: The {Finnish} Perspective",
  author   = "Tikkanen, Jani T and Kentt{\"a}, Tuomas and Porthan, Kimmo and
              Huikuri, Heikki V and Junttila, M Juhani",
  journal  = "Annals of Noninvasive Electrocardiology: the Official Journal of
              the International Society for Holter and Noninvasive
              Electrocardiology, Inc",
  volume   =  20,
  number   =  6,
  pages    = "526--533",
  month    =  nov,
  year     =  2015,
  language = "en"
}

@INPROCEEDINGS{Loshchilov_2019_AdamW,
  title     = "Decoupled Weight Decay Regularization",
  booktitle = "{ICLR}",
  author    = "Loshchilov, Ilya and Hutter, Frank",
  year      =  2019
}

@inproceedings{Ramprasaath_GRADCAM,
  author       = {Ramprasaath R. Selvaraju and
                  Michael Cogswell and
                  Abhishek Das and
                  Ramakrishna Vedantam and
                  Devi Parikh and
                  Dhruv Batra},
  title        = {{Grad-CAM}: Visual Explanations from Deep Networks via Gradient-Based
                  Localization},
  booktitle    = {{IEEE} International Conference on Computer Vision, {ICCV} 2017, Venice,
                  Italy, October 22-29, 2017},
  pages        = {618--626},
  publisher    = {{IEEE} Computer Society},
  year         = {2017},
  url          = {https://doi.org/10.1109/ICCV.2017.74},
  doi          = {10.1109/ICCV.2017.74},
  bibsource    = {dblp computer science bibliography, https://dblp.org}
}

@online{ecgwaves2023,
  title        = {{EKG/ECG} Leads, Electrodes, Limb Leads, Chest (Precordial) Leads},
  year         = 2023,
  url          = {https://ecgwaves.com/topic/ekg-ecg-leads-electrodes-systems-limb-chest-precordial/},
  note         = {Accessed on: 2023-07-01},
  organization = {ECG Waves},
}

@ARTICLE{Ilkhanoff_2012_EHF,
  title    = "Association of {QRS} duration with left ventricular structure and
              function and risk of heart failure in middle-aged and older
              adults: the {Multi-Ethnic} Study of Atherosclerosis ({MESA})",
  author   = "Ilkhanoff, Leonard and Liu, Kiang and Ning, Hongyan and Nazarian,
              Saman and Bluemke, David A and Soliman, Elsayed Z and
              Lloyd-Jones, Donald M",
  journal  = "European Journal of Heart Failure",
  volume   =  14,
  number   =  11,
  pages    = "1285--1292",
  month    =  nov,
  year     =  2012,
  language = "en"
}

@article{Turgut_2023_arxiv,
  title = {Unlocking the Diagnostic Potential of ECG through Knowledge Transfer from Cardiac MRI},
  author = "{Turgut, Ö. and Müller, P. and Hager, P. and Shit, S. and Starck, S. and Menten, M. J. and Martens, E. and Rueckert, D.}",
  journal = {ArXiv},
  year = {2023},
  url = {https://arxiv.org/abs/2308.05764}
}

@article{Strodthoff_2023_PTBXL,
  title     = {PTB-XL+, a comprehensive electrocardiographic feature dataset},
  author    = {Strodthoff, Nils and Mehari, Temesgen and Nagel, Claudia and Aston, Philip J. and Sundar, Ashish and Graff, Claus and Kanters, Jørgen K. and Haverkamp, Wilhelm and Dössel, Olaf and Loewe, Axel and Bär, Markus and Schaeffter, Tobias},
  journal   = {Scientific Data},
  volume    = {10},
  number    = {1},
  pages     = {279},
  year      = {2023},
  doi       = {10.1038/s41597-023-02153-8},
  pmid      = {37179420},
  issn      = {2052-4463},
}

@ARTICLE{Sudlow_2015_UKB,
  title    = "{UK} biobank: an open access resource for identifying the causes
              of a wide range of complex diseases of middle and old age",
  author   = "Sudlow, Cathie and Gallacher, John and Allen, Naomi and Beral,
              Valerie and Burton, Paul and Danesh, John and Downey, Paul and
              Elliott, Paul and Green, Jane and Landray, Martin and Liu, Bette
              and Matthews, Paul and Ong, Giok and Pell, Jill and Silman, Alan
              and Young, Alan and Sprosen, Tim and Peakman, Tim and Collins,
              Rory",
  journal  = "PLoS Medicine",
  volume   =  12,
  number   =  3,
  pages    = "e1001779",
  month    =  mar,
  year     =  2015,
  language = "en"
}

@ARTICLE{Zhang_2007_JACC_QRS,
  title    = "Comparison of the prognostic significance of the
              electrocardiographic {QRS/T} angles in predicting incident
              coronary heart disease and total mortality (from the
              atherosclerosis risk in communities study)",
  author   = "Zhang, Zhu-Ming and Prineas, Ronald J and Case, Douglas and
              Soliman, Elsayed Z and Rautaharju, Pentti M and {ARIC Research
              Group}",
  journal  = "The American journal of cardiology",
  volume   =  100,
  number   =  5,
  pages    = "844--849",
  month    =  sep,
  year     =  2007,
  language = "en"
}

@ARTICLE{Singhal_2022_Nature_LLM,
  title         = "Large Language Models Encode Clinical Knowledge",
  author        = "Singhal, Karan and Azizi, Shekoofeh and Tu, Tao and Sara
                   Mahdavi, S and Wei, Jason and Chung, Hyung Won and Scales,
                   Nathan and Tanwani, Ajay and Cole-Lewis, Heather and Pfohl,
                   Stephen and Payne, Perry and Seneviratne, Martin and Gamble,
                   Paul and Kelly, Chris and Scharli, Nathaneal and Chowdhery,
                   Aakanksha and Mansfield, Philip and Aguera y Arcas, Blaise
                   and Webster, Dale and Corrado, Greg S and Matias, Yossi and
                   Chou, Katherine and Gottweis, Juraj and Tomasev, Nenad and
                   Liu, Yun and Rajkomar, Alvin and Barral, Joelle and Semturs,
                   Christopher and Karthikesalingam, Alan and Natarajan, Vivek",
  month         =  dec,
  year          =  2022,
  journal  = {Nature},
  volume={620(7972)},
  pages = "172-180"
}

@inproceedings{Kingma_2013_VAE,
     title  = "{Auto-Encoding} Variational Bayes",
  author = "Kingma, Diederik P and Welling, Max",
  booktitle=ICLR,
  pages  = "1--14",
  year   =  2014
}

@article{zhang_2013_Circulation_different,
  title={Different patterns of bundle-branch blocks and the risk of incident heart failure in the {Women's Health Initiative (WHI)} study},
  author={Zhang, Zhe-Ming and Rautaharju, Pentti M and Soliman, Elsayed Z and Manson, JoAnn E and Martin, Lisa W and Perez, Marco and Vitolins, Mara and Prineas, Ronald J},
  journal={Circulation: Heart Failure},
  volume={6},
  pages={655--661},
  year={2013},
  publisher={Lippincott Williams \& Wilkins}
}

@article{rautaharju_2006_electrocardiographic,
  title={Electrocardiographic predictors of incident congestive heart failure and all-cause mortality in postmenopausal women: the Women's Health Initiative},
  author={Rautaharju, Pentti M and Kooperberg, Charles and Larson, Jennifer C and LaCroix, Andrea},
  journal={Circulation},
  volume={113},
  pages={481--489},
  year={2006},
  publisher={Lippincott Williams \& Wilkins}
}

@article{triola2005electrocardiographic,
  title={Electrocardiographic predictors of cardiovascular outcome in women: the National Heart, Lung, and Blood Institute-sponsored Women's Ischemia Syndrome Evaluation (WISE) study},
  author={Triola, Beth and Olson, Mary Beth and Reis, Steven E and Rautaharju, Pentti and Merz, C Noel Bairey and Kelsey, Sheryl F and Shaw, Leslee J and Sharaf, Barry L and Sopko, George and Saba, Samir},
  journal={Journal of the American College of Cardiology},
  volume={46},
  pages={51--56},
  year={2005},
  publisher={Elsevier}
}

@article{rautaharju2007electrocardiographic,
  title={Electrocardiographic predictors of new-onset heart failure in men and in women free of coronary heart disease (from the {Atherosclerosis in Communities [ARIC] Study})},
  author={Rautaharju, Pentti M and Prineas, Ronald J and Wood, John and Zhang, Zhe-Ming and Crow, Richard and Heiss, Gerardo},
  journal={The American Journal of Cardiology},
  volume={100},
  pages={1437--1441},
  year={2007},
  publisher={Elsevier}
}

@article{rautaharju2013electrocardiographic,
  title={Electrocardiographic predictors of incident heart failure in men and women free from manifest cardiovascular disease (from the Atherosclerosis Risk in Communities [ARIC] Study)},
  author={Rautaharju, Pentti M and Zhang, Zhe-Ming and Haisty Jr, William K and Prineas, Ronald J and Kucharska-Newton, Anna M and Rosamond, Wayne D and Soliman, Elsayed Z},
  journal={The American Journal of Cardiology},
  volume={112},
  pages={843--849},
  year={2013},
  publisher={Elsevier}
}

@ARTICLE{Zhang_2015_Ventricular_conduction,
  title    = "Ventricular conduction defects and the risk of incident heart
              failure in the Atherosclerosis Risk in Communities ({ARIC}) Study",
  author   = "Zhang, Zhu-Ming and Rautaharju, Pentti M and Prineas, Ronald J
              and Loehr, Laura and Rosamond, Wayne and Soliman, Elsayed Z",
  journal  = "Journal of Cardiac Failure",
  volume   =  21,
  number   =  4,
  pages    = "307--312",
  month    =  apr,
  year     =  2015,
  language = "en"
}

@ARTICLE{Silva_2013_P_wave,
  title    = "P-wave dispersion and left atrial volume index as predictors in
              heart failure",
  author   = "da Silva, Rose Mary Ferreira Lisboa and Kazzaz, Nadya Mendes and
              Torres, Ros{\'a}lia Morais and Moreira, Maria da Consola{\c
              c}{\~a}o Vieira",
  journal  = "Arquivos Brasileiros de Cardiologia",
  volume   =  100,
  number   =  1,
  pages    = "67--74",
  month    =  jan,
  year     =  2013,
}

@ARTICLE{Thompson_2004_JCF_Immediate,
  title     = "Immediate vs delayed diagnosis of heart failure: is there a
               difference in outcomes? results of a harris
               interactive\textregistered{} Patient Survey",
  author    = "Thompson, Brenda S and Yancy, Clyde W",
  journal   = "Journal of Cardiac Failure",
  publisher = "Elsevier",
  volume    =  10,
  number    =  4,
  pages     = "S125",
  month     =  aug,
  year      =  2004
}

@ARTICLE{Yancy_2006_JACC_Clinical_presentation,
  title    = "Clinical presentation, management, and in-hospital outcomes of
              patients admitted with acute decompensated heart failure with
              preserved systolic function: a report from the Acute
              Decompensated Heart Failure National Registry ({ADHERE}) Database",
  author   = "Yancy, Clyde W and Lopatin, Margarita and Stevenson, Lynne Warner
              and De Marco, Teresa and Fonarow, Gregg C and {ADHERE Scientific
              Advisory Committee and Investigators}",
  journal  = "Journal of the American College of Cardiology",
  volume   =  47,
  number   =  1,
  pages    = "76--84",
  month    =  jan,
  year     =  2006,
}

@ARTICLE{Khan_2022_Circulation,
  title    = "Development and Validation of a {Long-Term} Incident Heart
              Failure Risk Model",
  author   = "Khan, Sadiya S and Ning, Hongyan and Allen, Norrina B and
              Carnethon, Mercedes R and Yancy, Clyde W and Shah, Sanjiv J and
              Wilkins, John T and Tian, Lu and Lloyd-Jones, Donald M",
  journal  = "Circulation Research",
  volume   =  130,
  number   =  2,
  pages    = "200--209",
  month    =  jan,
  year     =  2022,
  language = "en"
}

@article{Khan_2019_JACC_10-Year,
author = {Sadiya S. Khan  and Hongyan Ning  and Sanjiv J. Shah  and Clyde W. Yancy  and Mercedes Carnethon  and Jarett D. Berry  and Robert J. Mentz  and Emily O’Brien  and Adolfo Correa  and Navin Suthahar  and Rudolf A. de Boer  and John T. Wilkins  and Donald M. Lloyd-Jones },
title = {10-Year Risk Equations for Incident Heart Failure in the General Population},
journal = {Journal of the American College of Cardiology},
volume = {73},
number = {19},
pages = {2388-2397},
year = {2019},
doi = {10.1016/j.jacc.2019.02.057},

URL = {https://www.jacc.org/doi/abs/10.1016/j.jacc.2019.02.057},
eprint = {https://www.jacc.org/doi/pdf/10.1016/j.jacc.2019.02.057}

}

@misc{bhf2019heartfailure,
    author = {{British Heart Foundation}},
    title = "Heart failure hospital admissions rise by a third in five years",
    year = {2019},
    month = {November},
    journal = {British Heart Foundation News Archive},
    url = {https://www.bhf.org.uk/what-we-do/news-from-the-bhf/news-archive/2019/november/heart-failure-hospital-admissions-rise-by-a-third-in-five-years},
    note = {(Accessed on Oct, 2023)}
}

@ARTICLE{Cikes_2019_EJHF_Machine,
  title    = "Machine learning-based phenogrouping in heart failure to identify
              responders to cardiac resynchronization therapy",
  author   = "Cikes, Maja and Sanchez-Martinez, Sergio and Claggett, Brian and
              Duchateau, Nicolas and Piella, Gemma and Butakoff, Constantine
              and Pouleur, Anne Catherine and Knappe, Dorit and
              Biering-S{\o}rensen, Tor and Kutyifa, Valentina and Moss, Arthur
              and Stein, Kenneth and Solomon, Scott D and Bijnens, Bart",
  journal  = "European Journal of Heart Failure",
  volume   =  21,
  number   =  1,
  pages    = "74--85",
  month    =  jan,
  year     =  2019,
  language = "en"
}

@article{cox1972regression,
  title={Regression models and life-tables},
  author={Cox, David R},
  journal={Journal of the Royal Statistical Society. Series B (Methodological)},
  volume={34},
  number={2},
  pages={187--220},
  year={1972},
  publisher={Wiley Online Library}
}

@ARTICLE{Katzman_2018_DeepSurv,
  title   = "{DeepSurv}: Personalized Treatment Recommender System Using A Cox
             Proportional Hazards Deep Neural Network",
  author  = "Katzman, Jared and Shaham, Uri and Bates, Jonathan and Cloninger,
             Alexander and Jiang, Tingting and Kluger, Yuval",
  journal = "BMC Medical Research Methodology",
  year    =  2018
}

@inproceedings{Liu2023_ICCV_CLIP,
  title={CLIP-Driven Universal Model for Organ Segmentation and Tumor Detection},
  author={Liu, Jie and Zhang, Yixiao and Chen, Jie-Neng and Xiao, Junfei and Lu, Yongyi and Landman, Bennett A and Yuan, Yixuan and Yuille, Alan and Tang, Yucheng and Zhou, Zongwei},
  booktitle=ICCV,
  year={2023}
}

@ARTICLE{Zhang_2023_Nature_communication,
  title    = "Knowledge-enhanced visual-language pre-training on chest
              radiology images",
  author   = "Zhang, Xiaoman and Wu, Chaoyi and Zhang, Ya and Xie, Weidi and
              Wang, Yanfeng",
  journal  = "Nature Communications",
  volume   =  14,
  number   =  1,
  pages    = "4542",
  month    =  jul,
  year     =  2023,
  language = "en"
}

@ARTICLE{Radford2021_CLIP,
  title         = "Learning Transferable Visual Models From Natural Language
                   Supervision",
  author        = "Radford, Alec and Kim, Jong Wook and Hallacy, Chris and
                   Ramesh, Aditya and Goh, Gabriel and Agarwal, Sandhini and
                   Sastry, Girish and Askell, Amanda and Mishkin, Pamela and
                   Clark, Jack and Krueger, Gretchen and Sutskever, Ilya",
  month         =  feb,
  year          =  2021,
  archivePrefix = "arXiv",
  primaryClass  = "cs.CV",
  eprint        = "2103.00020"
}

@inproceedings{Nejedly_2022_CINC_winner,
author={Nejedly, Petr and Ivora, Adam and Smisek, Radovan and Viscor, Ivo and Koscova, Zuzana and Jurak, Pavel and Plesinger, Filip},
  booktitle={2021 Computing in Cardiology (CinC)}, 
  title={Classification of ECG Using Ensemble of Residual CNNs with Attention Mechanism}, 
  year={2021},
  volume={48},
  number={},
  pages={1-4},
  doi={10.23919/CinC53138.2021.9662723}
}

@ARTICLE{Liu_2018_ICBEB2018,
  title   = "An Open Access Database for Evaluating the Algorithms of
             Electrocardiogram Rhythm and Morphology Abnormality Detection",
  author  = "Liu, Feifei and Liu, Chengyu and Zhao, Lina and Zhang, Xiangyu and
             Wu, Xiaoling and Xu, Xiaoyan and Liu, Yulin and Ma, Caiyun and
             Wei, Shoushui and He, Zhiqiang and Li, Jianqing and Yin Kwee,
             Eddie Ng",
  journal = "Journal of Medical Imaging and Health Informatics",
  volume  =  8,
  number  =  7,
  pages   = "1368--1373",
  year    =  2018
}

@misc{han_2023_brain,
      title={Can Brain Signals Reveal Inner Alignment with Human Languages?}, 
      author={William Han and Jielin Qiu and Jiacheng Zhu and Mengdi Xu and Douglas Weber and Bo Li and Ding Zhao},
      year={2023},
      eprint={2208.06348},
      archivePrefix={arXiv},
      primaryClass={q-bio.NC}
}

@ARTICLE{Liu_2023_ETP,
  title         = "{ETP}: Learning Transferable {ECG} Representations via
                   {ECG-Text} Pre-training",
  author        = "Liu, Che and Wan, Zhongwei and Cheng, Sibo and Zhang, Mi and
                   Arcucci, Rossella",
  month         =  sep,
  year          =  2023,
  archivePrefix = "arXiv",
  primaryClass  = "eess.SP",
  eprint        = "2309.07145"
}

@INPROCEEDINGS{Qiu_2023_EACL,
  title     = "Transfer Knowledge from Natural Language to Electrocardiography:
               Can We Detect Cardiovascular Disease Through Language Models?",
  booktitle = "Findings of the Association for Computational Linguistics:
               {EACL} 2023",
  author    = "Qiu, Jielin and Han, William and Zhu, Jiacheng and Xu, Mengdi
               and Rosenberg, Michael and Liu, Emerson and Weber, Douglas and
               Zhao, Ding",
  publisher = "Association for Computational Linguistics",
  pages     = "442--453",
  month     =  may,
  year      =  2023,
  address   = "Dubrovnik, Croatia"
}

@misc{Qiu_2023_Multimodal,
  title         = "Multimodal Representation Learning of Cardiovascular
                   Magnetic Resonance Imaging",
  author        = "Qiu, Jielin and Huang, Peide and Nakashima, Makiya and Lee,
                   Jaehyun and Zhu, Jiacheng and Tang, Wilson and Chen, Pohao
                   and Nguyen, Christopher and Kim, Byung-Hak and Kwon, Debbie
                   and Weber, Douglas and Zhao, Ding and Chen, David",
  month         =  apr,
  year          =  2023,
  archivePrefix = "arXiv",
  primaryClass  = "cs.CV",
  eprint        = "2304.07675"
}

@ARTICLE{Strodthoff_2021_Deep_Learning_for_PTB_XL,
  title    = "Deep Learning for {ECG} Analysis: Benchmarks and Insights from
              {PTB-XL}",
  author   = "Strodthoff, Nils and Wagner, Patrick and Schaeffter, Tobias and
              Samek, Wojciech",
  journal  = "IEEE Journal of Biomedical and Health Informatics",
  volume   =  25,
  number   =  5,
  pages    = "1519--1528",
  month    =  may,
  year     =  2021,
  language = "en"
}

@article{tomaselli2004causes,
  title={What causes sudden death in heart failure?},
  author={Tomaselli, Gordon F and Zipes, Douglas P},
  journal={Circulation Research},
  volume={95},
  number={8},
  pages={754--763},
  year={2004},
  publisher={Am Heart Assoc}
}

@article{lane_2005_prediction,
  title={Prediction and prevention of sudden cardiac death in heart failure},
  author={Lane, Rebecca E and Cowie, Martin R and Chow, Anthony WC},
  journal={Heart},
  volume={91},
  number={5},
  pages={674--680},
  year={2005},
  publisher={BMJ Publishing Group Ltd}
}

@ARTICLE{Wagner2020_PTB_XL,
  title    = "{PTB-XL}, a large publicly available electrocardiography dataset",
  author   = "Wagner, Patrick and Strodthoff, Nils and Bousseljot, Ralf-Dieter
              and Kreiseler, Dieter and Lunze, Fatima I and Samek, Wojciech and
              Schaeffter, Tobias",
  journal  = "Scientific Data",
  volume   =  7,
  number   =  1,
  pages    = "154",
  month    =  may,
  year     =  2020,
  language = "en"
}

@ARTICLE{Biton_2021_Atrial,
  title    = "Atrial fibrillation risk prediction from the 12-lead
              electrocardiogram using digital biomarkers and deep
              representation learning",
  author   = "Biton, Shany and Gendelman, Sheina and Ribeiro, Ant{\^o}nio H and
              Miana, Gabriela and Moreira, Carla and Ribeiro, Antonio Luiz P
              and Behar, Joachim A",
  journal  = "European Heart Journal. Digital health",
  volume   =  2,
  number   =  4,
  pages    = "576--585",
  month    =  dec,
  year     =  2021,
  language = "en"
}

@ARTICLE{McInnes2018_UMAP,
  title         = "{UMAP}: Uniform Manifold Approximation and Projection for
                   Dimension Reduction",
  author        = "McInnes, Leland and Healy, John and Melville, James",
  month         =  feb,
  year          =  2018,
  archivePrefix = "arXiv",
  primaryClass  = "stat.ML",
  eprint        = "1802.03426"
}

@ARTICLE{Ba2016_LayerNorm,
  title         = "Layer Normalization",
  author        = "Ba, Jimmy Lei and Kiros, Jamie Ryan and Hinton, Geoffrey E",
  month         =  jul,
  year          =  2016,
  archivePrefix = "arXiv",
  primaryClass  = "stat.ML",
  eprint        = "1607.06450"
}

@INPROCEEDINGS{Vaswani_2017_NIPS,
  title     = "Attention is All you Need",
  booktitle = NIPS,
  author    = "Vaswani, Ashish and Shazeer, Noam and Parmar, Niki and
               Uszkoreit, Jakob and Jones, Llion and Gomez, Aidan N and Kaiser,
               {\L} Ukasz and Polosukhin, Illia",
  editor    = "Guyon, I and Luxburg, U V and Bengio, S and Wallach, H and
               Fergus, R and Vishwanathan, S and Garnett, R",
  publisher = "Curran Associates, Inc.",
  pages     = "5998--6008",
  year      =  2017
}

@ARTICLE{Harrell1982_C_index,
  title    = "Evaluating the yield of medical tests",
  author   = "Harrell, Jr, F E and Califf, R M and Pryor, D B and Lee, K L and
              Rosati, R A",
  journal  = "JAMA: the Journal of the American Medical Association",
  volume   =  247,
  number   =  18,
  pages    = "2543--2546",
  month    =  may,
  year     =  1982,
  language = "en"
}

@INPROCEEDINGS{Lei_2022_STACOM,
  author       = {Lei Li and
                  Juli{\`{a}} Camps and
                  Abhirup Banerjee and
                  Marcel Beetz and
                  Blanca Rodr{\'{\i}}guez and
                  Vicente Grau},
  title        = {Deep Computational Model for the Inference of Ventricular Activation
                  Properties},
  booktitle    = STACOM,
  volume       = {13593},
  pages        = {369--380},
  publisher    = {Springer},
  year         = {2022},
  url          = {https://doi.org/10.1007/978-3-031-23443-9\_34},
  doi          = {10.1007/978-3-031-23443-9\_34},
  bibsource    = {dblp computer science bibliography, https://dblp.org}
}

@inproceedings{Marcel_2022_ISBI,
  author       = {Marcel Beetz and
                  Abhirup Banerjee and
                  Yuling Sang and
                  Vicente Grau},
  title        = {Combined Generation of Electrocardiogram and Cardiac Anatomy Models
                  Using Multi-Modal Variational Autoencoders},
  booktitle    = ISBI,
  pages        = {1--4},
  publisher    = {{IEEE}},
  year         = {2022},
  url          = {https://doi.org/10.1109/ISBI52829.2022.9761590},
  doi          = {10.1109/ISBI52829.2022.9761590},
  bibsource    = {dblp computer science bibliography, https://dblp.org}
}

@ARTICLE{Marcel_2022_Frontiers,
  AUTHOR={Beetz, Marcel and Banerjee, Abhirup and Grau, Vicente},    
TITLE={Multi-Domain Variational Autoencoders for Combined Modeling of {MRI}-Based Biventricular Anatomy and {ECG}-Based Cardiac Electrophysiology},  JOURNAL={Frontiers in Physiology},      
VOLUME={13},           	
YEAR={2022},      	  
URL={https://www.frontiersin.org/articles/10.3389/fphys.2022.886723},       
DOI={10.3389/fphys.2022.886723},      
ISSN={1664-042X}
}

@inproceedings{Yuling_2023,
  author       = {Yuling Sang and
                  Marcel Beetz and
                  Vicente Grau},
  title        = {Generation of 12-Lead Electrocardiogram with Subject-Specific, Image-Derived
                  Characteristics Using a Conditional Variational Autoencoder},
  booktitle    = ISBI,
  pages        = {1--5},
  publisher    = {{IEEE}},
  year         = {2022},
  url          = {https://doi.org/10.1109/ISBI52829.2022.9761431},
  doi          = {10.1109/ISBI52829.2022.9761431},
  bibsource    = {dblp computer science bibliography, https://dblp.org}
}

@ARTICLE{Li_2023_TMI_Survival,
  title    = "Survival Prediction via Hierarchical Multimodal {Co-Attention}
              Transformer: A Computational {Histology-Radiology} Solution",
  author   = "Li, Zhe and Jiang, Yuming and Lu, Mengkang and Li, Ruijiang and
              Xia, Yong",
  journal  = "IEEE Transactions on Medical Imaging",
  volume   = "PP",
  month    =  mar,
  year     =  2023,
  language = "en"
}

@ARTICLE{Jiang_2023_MH,
  title   = "{MHAttnSurv}: {Multi-Head} Attention for Survival Prediction Using
             {Whole-Slide} Pathology Images",
  author  = "Jiang, Shuai and Suriawinata, Arief A and Hassanpour, Saeed",
  journal = "Computers in Biology and Medicine",
  volume  =  158,
  month   =  may,
  year    =  2023
}

@ARTICLE{Bello_2019_NMI,
  title    = "Deep learning cardiac motion analysis for human survival
              prediction",
  author   = "Bello, Ghalib A and Dawes, Timothy J W and Duan, Jinming and
              Biffi, Carlo and de Marvao, Antonio and Howard, Luke S G E and
              Gibbs, J Simon R and Wilkins, Martin R and Cook, Stuart A and
              Rueckert, Daniel and O'Regan, Declan P",
  journal  = NMI,
  volume   =  1,
  pages    = "95--104",
  month    =  2,
  year     =  2019,
  issn     = "2522-5839",
  pmid     = "30801055",
  doi      = "10.1038/s42256-019-0019-2",
  pmc      = "PMC6382062"
}

\clearpage
\appendix
\setcounter{table}{0}
\renewcommand{\thetable}{A\arabic{table}}
\subsection{Implementation of temporal activation maps}
To visualize the temporal attention matrix $\mathbf{A}_i^{TA} \in \Reals^{l\times l} (l=\frac{T}{2^K}<T)$ in the original input space of the ECG with a time length of $T$, we employ a modified version of GradCAM (Gradient-weighted Class Activation Mapping)~\cite{Ramprasaath_GRADCAM}. This technique allows us to generate visualizations for each ECG lead covering the entire duration of the signal. Specifically, we start by adding up the values in each column of the attention matrix to get a single attention score for every moment along this reduced time dimension. Then, for each of these time points, we find the related input feature from the ECG and use it to create a gradient activation map, mapping back to the original input space. We repeat this process for every point in time and then combine all the resulting GradCAM maps. Each map is weighted according to its respective attention score, ensuring that moments with higher attention scores have a greater influence on the final visualization. 
\begin{table*}[!ht]
\centering
\caption{The code lists used for the retrieval of different disease records in UK Biobank database.}
\label{tab:ukb_code_retrieval}
\resizebox{\textwidth}{!}{%
\begin{tabular}{@{}lll@{}}
\toprule
Population                          & Category                               & Codes                                                                         \\ \midrule
\multirow{4}{*}{Heart Failure (HF)} & algorithmically defined HF & Field ID: 131354                                                              \\
                                    & self-report                            & Field ID: 20002, code: 1076                                                    \\
                                    & ICD 9                                  & Field ID: 41271, codes: 4280, 4281, 4289; Date field: 41281                   \\
                                    & ICD10                                  & Field ID: 41270, codes: I500, I501, I509, I110, I130, I132; Date field: 41280 \\ \midrule
\multirow{4}{*}{Myocardial Infarction (MI)} &
  algorithmically defined MI &
  Field ID: 131298 \\
                                    & self-report                            & Field ID: 20002, code: 1075                                                    \\
                                    & ICD 9                                  & Field ID: 41271, codes: 410, 411, 412, 436; Date field: 41281                    \\
                                    & ICD10                                  & Field ID: 41270, codes: I21, I22, I23, I24.1, I25.2; Date field: 41280        \\ \midrule
\multirow{4}{*}{Hypertension (HYP)} &
  algorithmically defined HYP &
  Field ID: 131286, 131288, 131289, 131292, 131293, 131294 \\
                                    & self-report                            & Field ID: 20002, code: 1065                                                    \\
                                    & ICD 9                                  & Field ID: 41271, codes: 4010, 4011, 4019; Date field: 41281                     \\
 &
  ICD 10 &
  \begin{tabular}[c]{@{}l@{}}Field ID: 41270, codes: I110, I13.0, I13.1, I13.2, I13.9, I15.1, I15.2, \\ I15.8, I15.9; Date field: 41280\end{tabular} \\ \bottomrule
\end{tabular}%
}
\end{table*}
\begin{table*}[]
\centering
\caption{Detailed configurations of the ECG dual attention-based encoder.}
\label{tab:encoder_structure}
\resizebox{\textwidth}{!}{%
\begin{tabular}{@{}lllll@{}}
\toprule
\multicolumn{5}{c}{\textbf{ECG signal dual attention encoder}} \\ \midrule
Layer Name &
  Input size &
  Output size &
  PyTorch Like structure &
  Description \\ \midrule
Conv1D &
  12$\times$1 $\times$ 1024 &
  12$\times$4 $\times$ 1024 &
  \begin{tabular}[c]{@{}l@{}}Conv1d( in\_ch= 12, out\_ch= 12$\times$4, ks=5, stride=1, p=2,groups=12),\\ GroupNorm(12,12$\times$4),\\ nn.GELU(),\\ Conv1d( in\_ch= 12$\times$4, out\_ch= 12$\times$4, ks=5, stride=1, p=2,groups=12),\\ GroupNorm(12,12$\times$4),\\ nn.GELU(),\end{tabular} &
  \begin{tabular}[c]{@{}l@{}}using group conv and group normal so that each lead \\ has a separate set of filters for signal preprocessing \\ and feature extraction\end{tabular} \\ \midrule
Reshape &
  12$\times$4$\times$1024 &
  4$\times$ 12$\times$ 1024 &
  Reshape &
  reshape it to a matrix (n\_feature, n\_leads, n\_length) \\ \midrule
 &
  4$\times$ 12$\times$ 1024 &
  8$\times$ 12$\times$ 512 &
  ResBlock(in\_ch=4, out\_ch=8, ks=5, stride=1, p=2, downsample= 2) &
   \\
 &
  8$\times$ 12$\times$ 512 &
  16$\times$ 12$\times$ 256 &
  ResBlock(in\_ch=8, out\_ch=16, ks=5, stride=1, p=2,downsample= 2) &
   \\
 &
  16$\times$12$\times$ 256 &
  16$\times$12$\times$128 &
  ResBlock(in\_ch=16, out\_ch=16, ks=5, stride=1, p=2,downsample= 2) &
   \\
 &
  16$\times$12$\times$128 &
  16$\times$12$\times$64 &
  ResBlock(in\_ch=16, out\_ch=16, ks=5, stride=1, p=2,downsample= 2) &
   \\
\multirow{-5}{*}{\begin{tabular}[c]{@{}l@{}}ResConv1D \\ (K=5; downsample)\end{tabular}} &
  16$\times$12$\times$ 64 &
  16$\times$12$\times$ 32 &
  ResBlock(in\_ch=16, out\_ch=16, ks=5, stride=1, p=2, downsample= 2) &
  \multirow{-5}{*}{\begin{tabular}[c]{@{}l@{}}peform feature extraction along the time dimension (the last) \\ while keeping the lead dimension unchanged\end{tabular}} \\ \midrule
DualAttention &
  16$\times$12$\times$ 32 &
  16$\times$12$\times$ 32 &
  LeadAttention(input = 12$\times${[}16$\times$32{]}) +Concat(12-lead TimeAttention (input = 32$\times$16)) &
  dual attention module \\ \midrule
Max-pooling &
  {[}16x12{]}$\times$32 &
  256 &
  \begin{tabular}[c]{@{}l@{}}Dropout(p=0.2);\\ Conv1d(in\_ch= 16$\times$12, out\_ch= 256, ks=1, stride=1, p=0);\\ AdaptiveMaxPool(output\_size=1):\\ Flatten()\end{tabular} &
  feature dimension reduction \\ \midrule
AVG-pooling &
  {[}16$\times$12{]}$\times$32 &
  256 &
 \begin{tabular}[c]{@{}l@{}}Dropout(p=0.2); \\ Conv1d(in\_ch= 16$\times$12, out\_ch= 256, ks=1, stride=1, p=0); \\ AdaptiveAVGPool(output\_size=1);\\ Flatten()\end{tabular} &
  feature dimension reduction \\ \midrule
Concat &
  256; 256 &
  512 &
  Concat &
  Concat features from Max-and AVG pooling \\ \bottomrule
\end{tabular}%
}
\end{table*}
\begin{table*}[]
\centering
\caption{Detailed configurations of the ECG decoder for signal reconstruction.}
\label{tab:decoder_architecture}
\resizebox{\textwidth}{!}{%
\begin{tabular}{@{}lllll@{}}
\toprule
\multicolumn{5}{c}{\textbf{Signal decoder}}                                                                                                                           \\ \midrule
\textbf{Layer Name} & \textbf{Input size}     & \textbf{Output size} & \textbf{PyTorch like structure}                                 & \textbf{Description}         \\ \midrule
Linear              & \multicolumn{1}{r}{512} & 4$\times$12$\times$ 32             & Linear (512, 12$\times$4$\times$32)                                           & feature refine and expansion \\ \midrule
\multirow{5}{*}{\begin{tabular}[c]{@{}l@{}}ResConv1D\\  (K=5; upsample)\end{tabular}} &
  4$\times$12$\times$32 &
  4$\times$12$\times$64 &
  ResBlock(in\_ch=4, out\_ch=4, ks=5, stride=1, p=2, upsample= 2) &
  \multirow{5}{*}{feature expansion along the time dimension} \\
                    & 4$\times$12$\times$64                & 4$\times$12$\times$128            & ResBlock(in\_ch=4, out\_ch=4, ks=5, stride=1, p=2, upsample= 2) &                              \\
                    & 4$\times$12$\times$ 128               & 4$\times$12$\times$ 256            & ResBlock(in\_ch=4, out\_ch=4, ks=5, stride=1, p=2, upsample= 2) &                              \\
                    & 4$\times$12$\times$ 256               & 4$\times$12$\times$ 512            & ResBlock(in\_ch=4, out\_ch=4, ks=5, stride=1, p=2, upsample= 2) &                              \\
                    & 4$\times$12$\times$ 512               & 1$\times$12$\times$ 1024           & ResBlock(in\_ch=4, out\_ch=1, ks=5, stride=1, p=2, upsample= 2) &                              \\ \midrule
Conv1D &
  12$\times$1$\times$1024 &
  12$\times$1$\times$1024 &
  \begin{tabular}[c]{@{}l@{}}Conv1d( in\_ch= 1, out\_ch=1, ks=1, stride=1, p=0,groups=12)\\ InstanceNorm1d(12)\end{tabular} &
  smooth and signal normalization \\ \bottomrule
\end{tabular}%
}
\end{table*}
\begin{table}[]
\centering
\caption{Detailed configurations of the risk prediction branch.}
\label{tab:risk_prediction_branch}
\resizebox{\columnwidth}{!}{%
\begin{tabular}{@{}lllll@{}}
\toprule
\multicolumn{5}{c}{\textbf{Risk prediction branch}}                          \\ \midrule
Layer Name & Input size & Output size & PyTorch Like structure & Description \\ \midrule
Linear          & 512 & 3 & \begin{tabular}[c]{@{}l@{}}BatchNorm (512);\\ BatchwiseDropout (0.25);\\ Linear(512,3);\\ ReLU;\end{tabular} & dimension reduction      \\
Risk prediction & 3   & 1 & \begin{tabular}[c]{@{}l@{}}BatchNorm (3);\\ Linear(3,1)\end{tabular}                                         & regression to risk score \\ \bottomrule
\end{tabular}%
}
\end{table}
\begin{table}[]
\centering
\caption{Detailed configurations of the two feature projectors.}
\label{tab:projector_architecture}
\resizebox{\columnwidth}{!}{%
\begin{tabular}{@{}cllll@{}}
\toprule
\multicolumn{5}{c}{\textbf{ECG projector $p_x$}} \\ \midrule
\textbf{Layer Name} &
  \multicolumn{1}{c}{\textbf{Input size}} &
  \multicolumn{1}{c}{\textbf{Output size}} &
  \multicolumn{1}{c}{\textbf{PyTorch Like structure}} &
  \textbf{Description} \\ \midrule
\multicolumn{1}{l}{Projection} &
  512 &
  128 &
  \begin{tabular}[c]{@{}l@{}}Linear(512, 256);\\ ReLU;\\ Linear (256, 128)\end{tabular} &
  feature reduction for feature alignment \\ \midrule
\multicolumn{5}{c}{\textbf{Text projector} $p_y$} \\ \midrule
\textbf{Layer Name} &
  \multicolumn{1}{c}{\textbf{Input size}} &
  \multicolumn{1}{c}{\textbf{Output size}} &
  \multicolumn{1}{c}{\textbf{PyTorch like structure}} &
  \textbf{Description} \\ \midrule
\multicolumn{1}{l}{Projection} &
  768 &
  128 &
  \begin{tabular}[c]{@{}l@{}}Linear(768, 256);\\ ReLU;\\ Linear (256, 128)\end{tabular} &
  feature reduction for feature alignment \\ \bottomrule
\end{tabular}%
}
\end{table}

\end{document}